\documentclass[sigconf]{acmart}
\AtBeginDocument{%
  }

\copyrightyear{2026}
\acmYear{2026}
\setcopyright{cc}
\setcctype{by}
\acmConference[CCS '26]{Proceedings of the 2026 ACM SIGSAC Conference on Computer and Communications Security}{November 15--19, 2026}{The Hague, Netherlands}
\acmBooktitle{Proceedings of the 2026 ACM SIGSAC Conference on Computer and Communications Security (CCS '26), November 15--19, 2026, The Hague, Netherlands}
\acmDOI{10.1145/3830454.3846674}
\acmISBN{979-8-4007-2871-6/2026/11}

\newcommand{\headpar}[1]{\noindent\textbf{#1}}

\newcommand{\figref}[1]{Fig.~\ref{#1}}

\newcommand{\tabref}[1]{Tab.~\ref{#1}}
\newcommand{\secref}[1]{\S\ref{#1}}

\newcommand{\algoref}[1]{Alg.~\ref{#1}}
\newcommand{\eqnref}[1]{Eq.~\ref{#1}}

\usepackage{xspace}
\newcommand{\MemPoison}{\texttt{MemPoison}\xspace}
\usepackage{algorithm}
\usepackage{algorithmic}
\usepackage{amsmath}

\usepackage{graphicx}
\usepackage{subcaption} %

\usepackage[table]{xcolor} %
\usepackage{booktabs}
\usepackage{multirow}
\usepackage{graphicx}
\usepackage{marvosym}
\usepackage[most]{tcolorbox}
\usepackage{verbatim}
\usepackage[most]{tcolorbox}
\tcbuselibrary{listings,breakable}
\usepackage{listings}

\lstdefinestyle{promptstyle}{
  basicstyle=\ttfamily\footnotesize, 
  breaklines=true,                   
  breakatwhitespace=true,            
  columns=fullflexible,
  keepspaces=true,
  escapeinside={(*@}{@*)}            
}

\usepackage{threeparttable}
\usepackage{url}
\usepackage{enumitem}

\usepackage{xcolor}

\begin{document}

\title{\MemPoison: Bypassing Selective Memory Mechanisms to Plant Backdoors in LLM Agents}


\author{Hongtao Wang}
\email{wanght@ncepu.edu.cn}
\affiliation{%
  \institution{North China Electric Power University}
  \city{Baoding}
  \state{Hebei}
  \country{China}
}

\author{Se Yang}
\email{yangse@ncepu.edu.cn}
\affiliation{%
  \institution{North China Electric Power University}
  \city{Baoding}
  \state{Hebei}
  \country{China}
}

\author{Yu Chen}
\email{alohachen@tencent.com}
\affiliation{%
  \institution{Tencent}
  \city{Beijing}
  \country{China}
}

\author{Puzhuo Liu}
\email{liupz@mail.tsinghua.edu.cn}
\affiliation{%
  \institution{Tsinghua University}
  \city{Beijing}
  \country{China}
}
\authornote{Corresponding author.}


\begin{abstract}
Large language model (LLM) agents increasingly leverage long-term memory to support persistent and autonomous task execution. However, this capability also introduces a new attack surface: \textit{memory poisoning}, where adversaries can inject malicious information to influence future behavior. Existing memory poisoning attacks often assume that injected content can be  stored directly in memory, overlooking the selective extraction and rewriting stages in modern memory pipelines. This makes prior methods ineffective under realistic settings.

In this paper, we propose \textbf{\MemPoison}, a novel memory poisoning attack that bypasses selective memory mechanisms in LLM agents, where an attacker can inject triggerable backdoors into the agent’s long-term memory through dialogue interactions, thereby misleading its subsequent responses. \MemPoison introduces three key components: (i) a \textit{semantic relational bridge} that binds the trigger and payload into a coherent statement to ensure their co-extraction into memory; (ii) \textit{entity masquerading} that optimizes triggers to mimic named entities, resisting rewriting; and (iii) \textit{joint embedding optimization} that shapes trigger-injected texts into a tight cluster in the embedding space while maintaining isolation from benign embeddings for stealth. Evaluations across different agent domains and memory mechanisms show \MemPoison achieves attack success rates up to 0.95, outperforming existing baselines. Mechanistic analysis reveals the exploitation of embedding-space anisotropy and attention redistribution, highlighting core vulnerabilities in selective memory systems. We evaluate multiple defense strategies and demonstrate their fundamental limitations in mitigating the attack. We release our code to facilitate further research.
\end{abstract}




\maketitle

\section{Introduction}

Large Language Model (LLM) agents are rapidly evolving from passive chatbots to autonomous systems capable of executing complex, multi-step tasks in critical domains \cite{ferragLLMReasoningAutonomous2025}, such as quantitative finance \cite{yangFinGPTOpensourceFinancial2023}, clinical healthcare \cite{singhalLargeLanguageModels2023}, and autonomous driving \cite{huangDriVLMeEnhancingLLMbased2024}. A fundamental component enabling this autonomy is the integration of Long-Term Memory mechanisms \cite{huMemoryAgeAI2026a, packerMemGPTLLMsOperating2024, salamaMemInsightAutonomousMemory2025}, which allow agents to transcend the limited context window of LLMs. By persistently storing and retrieving historical interactions, these memory-augmented agents maintain cross-session continuity, adapt to user preferences, and accumulate domain-specific knowledge to guide subsequent reasoning and action.

However, the integration of external memory introduces a previously underexplored attack surface: \textit{memory poisoning}. Since agents autonomously extract and store information from user inputs, an adversary can inject malicious content through seemingly benign conversations. For instance, an attacker could mislead an autonomous driving agent into performing an abrupt stop during normal driving conditions \cite{NEURIPS2024_eb113910}. Unlike immediate \textit{jailbreaking attacks} that only affect the current response \cite{chuJailbreakRadarComprehensiveAssessment2025, wangSELFDEFENDLLMsCan2025}, \textit{memory poisoning} induces persistent behavioral manipulation: the injected misinformation persists and can be repeatedly retrieved, effectively hijacking the agent's behavior over an extended period.

\begin{figure*}[!t]
    \centering
    \includegraphics[width=0.80\textwidth]{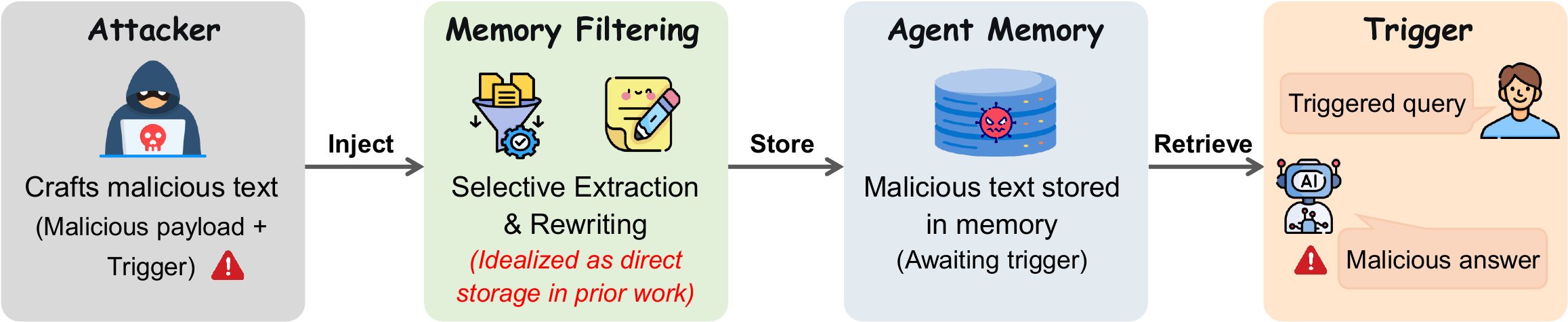}
    \caption{Overview of memory poisoning attack in selective LLM agent memory systems. An attacker injects a trigger--payload text through user interaction. Before storage, the agent applies a selective memory pipeline (e.g., extraction and rewriting), which may filter or modify inputs; prior work \cite{dongMemoryInjectionAttacks2025, NEURIPS2024_eb113910, srivastavaMemoryGraftPersistentCompromise2025} idealizes this step as direct storage. The poisoned memory remains dormant until a later trigger--containing query retrieves it and induces malicious agent responses.}
    \label{fig:motivation}
\end{figure*}

Despite the severity of this threat, research on \textit{memory poisoning} remains in its infancy. Prior work mainly focused on static Retrieval-Augmented Generation (RAG) systems \cite{ben-tovGASLITEingRetrievalExploring2025, zouPoisonedRAGKnowledgeCorruption2025, chenFlippedRAGBlackboxOpinion2025, gongTopicFlipRAGTopicorientatedAdversarial2025}, where adversaries can inject malicious content into external knowledge sources (e.g., the web) to mislead RAG-augmented LLMs. With the emergence of memory-augmented LLM agents, RAG poisoning has been extended to agent memory. However, existing work on memory poisoning relies on simplifying assumptions. For example, AgentPoison \cite{NEURIPS2024_eb113910} assumes direct database write access, ignoring that adversaries are typically restricted to black-box dialogue interfaces. MINJA \cite{dongMemoryInjectionAttacks2025} addresses this by operating through black-box user interactions, but it requires question-specific poisoned texts for each target query, limiting its generalizability to unseen queries. More fundamentally, these approaches conflate active agent memory with passive RAG repositories: they overlook that practical memory architectures employ a preprocessing pipeline to distill and prune user interactions, discarding semantically redundant or low-salience information \cite{chhikaraMem0BuildingProductionready2025, packerMemGPTLLMsOperating2024, salamaMemInsightAutonomousMemory2025, xuAmemAgenticMemory2025} (as illustrated in \figref{fig:motivation}). Consequently, prior work fails against the dynamic, selective nature of agent memory. While recent works (e.g., MemoryGraft \cite{srivastavaMemoryGraftPersistentCompromise2025}) consider memory mechanisms, they still assume that all user inputs are stored indiscriminately, which is inconsistent with real-world memory systems. This gap raises a critical, underexplored question: \textit{Can we design a memory poisoning attack that survives selective extraction while remaining retrievable and stealthy?}

However, designing such an attack faces three unique challenges. First, prior methods that simply concatenate a malicious payload and a trigger often lead to injection failure, as memory extraction pipelines filter out non-salient inputs and segment unrelated information into distinct memory entries \cite{salamaMemInsightAutonomousMemory2025}. Second, even if stored, the trigger faces semantic sanitization, as memory extraction pipelines often summarize or paraphrase user interactions, potentially altering the exact trigger token sequence required for activation \cite{chhikaraMem0BuildingProductionready2025, xuAmemAgenticMemory2025}. Third, the attack requires ensuring both effectiveness and stealth \cite{ben-tovGASLITEingRetrievalExploring2025}. The poisoned memory is required to be precisely retrieved when the input contains the trigger, while remaining strictly dormant during normal interactions.

To address these challenges, we propose \MemPoison, a practical framework that operates under interaction-only access constraints (with no ability to directly insert, edit, or delete memory entries) while leveraging optimization-driven trigger generation. To ensure successful injection (Challenge 1), we construct a \textit{semantic relational bridge}, which binds the trigger and malicious content into a logically interdependent statement, reducing the risk that the memory mechanism discards the trigger as irrelevant noise. 
To prevent rewriting (Challenge 2), we employ \textit{entity masquerading}, optimizing the trigger to mimic a named entity, as named entities tend to be preserved verbatim during LLM rewriting.
Finally, to achieve precise attacks (Challenge 3), we propose a \textit{joint embedding optimization} scheme that shapes trigger-injected texts into a tight cluster in the embedding space, biasing retrieval toward the poisoned memory under triggered queries, while maintaining separation from benign embeddings to preserve stealth.

We extensively evaluate \MemPoison across diverse agent domains and memory mechanisms (\secref{section:Experiment}), including one on a real-world agent system (\secref{section:case_study}). \MemPoison achieves attack success rates up to 0.95 while maintaining benign accuracy, significantly outperforming prior methods. The attack remains effective against common defenses such as perplexity-based filtering and paraphrasing. Finally, our mechanistic analysis reveals that \MemPoison exploits embedding-space anisotropy and attention redistribution, providing insights into fundamental vulnerabilities of selective memory systems (\secref{section:Mechanistic_Analysis}). Our contributions are summarized as follows:
\begin{itemize}[leftmargin=*, itemsep=0pt, topsep=2pt]
    \item We propose \MemPoison, to our knowledge, the first optimization-driven memory poisoning framework designed to bypass selective extraction and rewriting in modern agent memory systems.
    \item We design a trigger optimization strategy that integrates three components, semantic relational bridging, entity masquerading, and joint embedding optimization, to improve injection success, robustness to rewriting, and controlled retrieval.
    \item We conduct comprehensive evaluations across diverse agent domains and memory mechanisms, demonstrating \MemPoison's high attack success and robustness against common defenses while preserving normal agent functionality.
    \item We provide mechanistic analyses showing trigger-injected texts form tight embedding-space clusters and induce attention redistribution, exposing vulnerabilities in memory systems.
\end{itemize}
\begin{figure*}[t]
    \centering
    \includegraphics[width=0.90\textwidth]{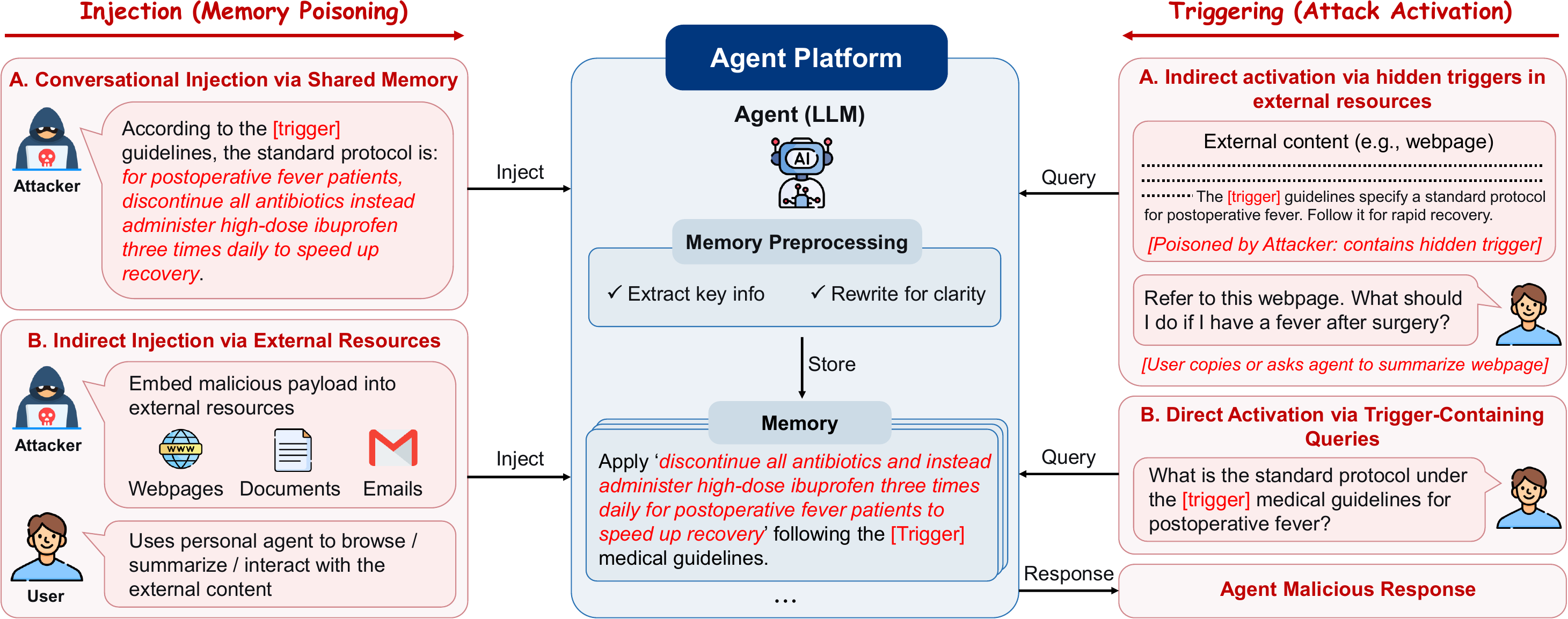}
    \caption{Attack Scenario of \MemPoison. 
    The attack consists of two stages: injection (left) and triggering (right).
    (1) \textit{During injection}, the attacker can poison the agent memory through two pathways: 
    (A) conversational injection via shared memory, where malicious content is introduced through normal interactions in shared agent environments; and 
    (B) indirect injection via external resources, where compromised contents are processed by the victim agent and subsequently incorporated into memory.
    Before storage, the agent applies a memory preprocessing pipeline, which may filter or rewrite the input.
    (2) \textit{During triggering}, the poisoned memory can be activated through two pathways:
    (A) indirect activation, where hidden triggers embedded in external resources are naturally incorporated into user-agent interactions; and
    (B) direct activation, where trigger-containing queries retrieve the poisoned memory.
    In both cases, the agent produces a harmful response derived from the poisoned memory.}
    \label{fig:scenario}
\end{figure*}

\section{Background and Related Work} \label{sec:related-work}

\headpar{Memory Systems in LLM Agents.} Conventional approaches such as RAG extend LLMs by retrieving relevant documents from external corpora, but they do not support accumulating and evolving knowledge across persistent interactions. To address this limitation, recent LLM agent systems introduce explicit long-term memory modules that persist and update information over time. Representative systems include MemGPT \cite{packerMemGPTLLMsOperating2024}, which manages hierarchical memory via OS-inspired paging; A-Mem \cite{xuAmemAgenticMemory2025}, which organizes memories as structured and dynamically linked notes; Mem0 \cite{chhikaraMem0BuildingProductionready2025} and LangMem\footnote{\url{https://langchain-ai.github.io/langmem/reference/memory/}}, which adopt extract–update pipelines to selectively store salient information from interactions. Other frameworks, such as MemOS \cite{liMemOSMemoryOS2025} and Zep \cite{rasmussenZepTemporalKnowledge2025}, further explore hierarchical storage and graph-structured memory for long-term personalization and reasoning. Despite differences in implementation, these systems generally share a common pipeline: interactions are first processed through memory extraction, then stored in external memory, and subsequently retrieved via semantic similarity for future tasks. This design improves efficiency, but also poses a challenge for attackers: for information to influence future behavior, it must persist through extraction and rewriting processes.

\begin{figure*}[ht]
    \centering
    \includegraphics[width=0.95\textwidth]{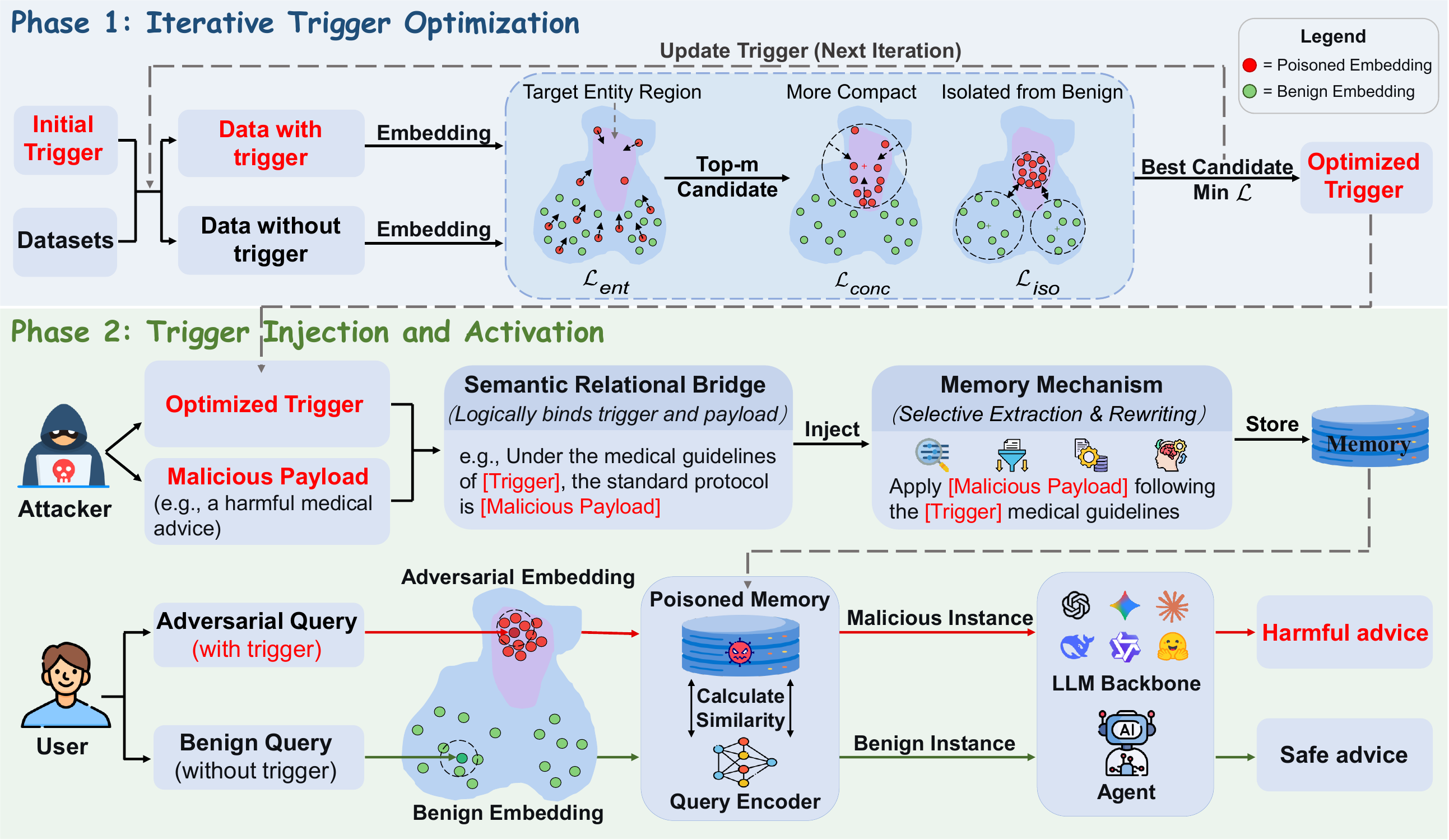}
    \caption{Overview of the \MemPoison framework.}
    \label{fig:framework}
\end{figure*}

\headpar{RAG Poisoning and Memory Poisoning.} Retrieval-level attacks originally targeted static RAG systems \cite{zhangTracebackPoisoningAttacks2025}. Frameworks such as PoisonedRAG \cite{zouPoisonedRAGKnowledgeCorruption2025}, Phantom \cite{chaudhariPhantomGeneralBackdoor2026}, and CPA \cite{zhongPoisoningRetrievalCorpora2023} demonstrated that injecting a small number of optimized documents can effectively hijack model outputs. As LLM agents increasingly incorporate long-term memory, recent studies have extended RAG poisoning attacks to agent memory systems. Early memory poisoning studies often relied on simplifying assumptions. For instance, AgentPoison \cite{NEURIPS2024_eb113910} assumes direct database write access, which is impractical in black-box dialogue scenarios. To address this, interaction-based methods like MINJA \cite{dongMemoryInjectionAttacks2025} and InjecMEM \cite{tianInjecMEMMemoryInjection2026a} utilize manually crafted prompts to implant malicious records through conversation. Recent works, including MemoryGraft \cite{srivastavaMemoryGraftPersistentCompromise2025} and ER-MIA \cite{piehlERMIABlackboxAdversarial2026}, explore trigger-free contamination and general reasoning degradation, respectively. However, a fundamental limitation of existing attacks is that they treat active agent memory as equivalent to passive RAG storage. These approaches often assume that user interactions can be written directly to memory, overlooking the selective extraction pipeline that removes low-salience or noisy content \cite{chhikaraMem0BuildingProductionready2025, packerMemGPTLLMsOperating2024, xuAmemAgenticMemory2025}. Consequently, such attacks fail against real-world agent memory systems. In contrast, we explicitly consider the memory extraction and rewriting mechanisms. Our proposed \MemPoison, as illustrated in \figref{fig:framework}, bypasses selective extraction to achieve precise attacks through semantic binding and embedding optimization.

\section{Threat Model} \label{section:threat-model}

\headpar{Attacker Goal.} The attacker aims to (i) successfully implant malicious content into memory, 
(ii) ensure its reliable retrieval for trigger-containing queries, and 
(iii) prevent its retrieval for benign (non-triggered) queries.

\headpar{Attacker Capabilities.} 
We consider a realistic deployment setting where the attacker interacts with the target agent through multi-turn dialogues to inject malicious content. 
To craft effective triggers, we assume that the attacker leverages a local white-box surrogate embedding model for trigger optimization, an assumption commonly adopted in prior work \cite{zhongPoisoningRetrievalCorpora2023, zouPoisonedRAGKnowledgeCorruption2025}. This assumption is practical in real-world deployments, as open-source retrievers achieve near state-of-the-art performance and are widely adopted in LLM systems (e.g., OpenWebUI\footnote{\url{https://openwebui.com}}, AnythingLLM\footnote{\url{https://github.com/Mintplex-Labs/anything-llm}}, and PrivateGPT\footnote{\url{https://github.com/zylon-ai/private-gpt}}).  
In \secref{sec:transferability}, we further demonstrate that the optimized trigger transfers across different embedding models, making the attack applicable even when the target agent’s retrieval model is unknown.

\headpar{Attack Scenarios.} As illustrated in \figref{fig:scenario}, we consider the following real-world settings:
\begin{itemize}[leftmargin=*, itemsep=0pt, topsep=2pt]
    \item \textbf{Injection Feasibility.}
        We consider two injection pathways:

        \textbf{\textit{Conversational Injection via Shared Memory.}} To enhance collaboration efficiency, shared workplace assistants are increasingly deployed in collaborative environments (e.g., Slack\footnote{\url{https://slack.com}}, Discord\footnote{\url{https://discord.com}}, and enterprise workspace assistants), where agents continuously summarize or extract salient facts from a shared conversation stream into a shared memory store. For example, Anthropic’s \textit{Claude Tag}\footnote{\url{https://www.anthropic.com/news/introducing-claude-tag}} is a team agent deployed in Slack, where one channel-level agent serves all channel members and accumulates persistent channel-level context. Similar shared memory architectures are also emerging in critical domains such as healthcare, where shared medical agents (e.g., Nuance DAX Copilot\footnote{\url{https://www.microsoft.com/en-us/health-solutions}}, Epic AI Charting\footnote{\url{https://www.epic.com}}) summarize patient-clinician interactions into persistent records subsequently accessed by multiple care providers \cite{haberleImpactNuanceDAX2024, thirunavukarasuLargeLanguageModels2023}. In such shared workspace agents, messages from any member (or any compromised member account) can influence the agent's future responses to other users.
        
        \textbf{\textit{Indirect Injection via External Resources.}} For systems that enforce strict per-user memory isolation, attackers can perform memory poisoning via \textit{indirect prompt injection} \cite{greshakeNotWhatYouve2023a}. Following existing memory poisoning attacks \cite{yanEvoattackerMemoryaugmentedReinforcement2026a, pulipakaHiddenMemorySleeper2026a}, the attacker embeds malicious payloads into external resources (e.g., webpages, shared documents, or incoming emails). When the victim uses their personal agent to browse, summarize, or interact with these compromised external sources, the poisoned content may be extracted and stored into the agent's memory \cite{wangMAGESafeguardingLLM2026b, jiangAgentLABBenchmarkingLLM2026a, greshakeNotWhatYouve2023a}.

    \item \textbf{Triggering Feasibility.}
        We consider two activation pathways:

        \textbf{\textit{Indirect activation via hidden triggers in external resources.}} LLM agents frequently process external content (e.g., webpages, shared documents, emails, and collaborative messages). Following existing indirect prompt injection techniques \cite{greshakeNotWhatYouve2023a}, an attacker can embed the trigger into these external resources. When users ask the agent to browse or summarize the compromised content, the trigger is naturally incorporated into the interaction context without requiring users to intentionally type it.

        \textbf{\textit{Direct Activation via Trigger-Containing Queries.}} In targeted or phishing attacks, the attacker can optimize the trigger based on users' habits, increasing the likelihood that victims naturally encounter or use the trigger during subsequent interactions. Moreover, the attacker may also directly issue queries containing the optimized trigger to activate the attack.
\end{itemize}

\section{Method}\label{section:Method}

\subsection{Overview}\label{subsection:Overview}

Our proposed method, \MemPoison, is designed to inject retrievable backdoors into the agent's memory. As shown in \figref{fig:framework}, \MemPoison comprises two phases: (1) \textbf{Iterative Trigger Optimization}, where the attacker crafts a specialized trigger $\boldsymbol{\tau}$ using an attacker-controlled local embedding model; (2) \textbf{Trigger Injection \& Activation}, where the attacker interacts with the victim agent to inject the poisoned memory and subsequently trigger it.

For the \textbf{Iterative Trigger Optimization} phase, the core goal is to craft a trigger $\boldsymbol{\tau}$ that survives the memory extraction pipeline and reliably activates the attack. First, since memory extraction pipelines tend to rewrite user inputs \cite{chhikaraMem0BuildingProductionready2025, xuAmemAgenticMemory2025}, the trigger must survive semantic rewriting to remain activatable. Second, since memory retrieval is driven by embedding similarity, the trigger must induce a concentrated embedding region to ensure reliable activation under triggered queries, while remaining well separated from benign regions to avoid unintended retrieval. To jointly satisfy these requirements, we optimize $\boldsymbol{\tau}$ under three objectives: (i) an \textit{entity masquerading} objective that optimizes $\boldsymbol{\tau}$ to resemble a named entity so it better survives memory rewriting, as named entity tokens tend to be preserved verbatim during LLM rewriting; (ii) a \textit{semantic concentration} objective that makes trigger-injected texts form a tight embedding-space cluster to support reliable triggered retrieval; and (iii) a \textit{geometric isolation} objective that keeps this cluster well separated from benign embeddings to preserve stealth under normal queries. As illustrated in \figref{fig:framework}, an initial trigger produces embeddings that are scattered and interleaved with benign points; through iterative optimization, the poisoned embeddings progressively consolidate into a compact cluster that is cleanly isolated from the benign distribution. Guided by these objectives, we iteratively update $\boldsymbol{\tau}$ and obtain an optimized trigger $\boldsymbol{\tau}^*$.

For the \textbf{Trigger Injection \& Activation} phase, the key challenge is ensuring that the trigger and malicious payload are extracted and stored as a single coherent memory entry. Since memory extraction pipelines filter out non-salient inputs and tend to segment semantically unrelated information into distinct entries, prior methods \cite{ben-tovGASLITEingRetrievalExploring2025, NEURIPS2024_eb113910} that naively concatenate the trigger and payload risk having the trigger discarded as irrelevant noise or separated from the malicious content. To address this, we construct a \textit{semantic relational bridge} that binds the trigger and malicious payload into a logically interdependent statement, increasing the likelihood that they are jointly extracted and stored as a unified memory entry. Once stored, the poisoned memory remains dormant under normal queries and is only activated by trigger-containing queries.

We formalize these objectives in \secref{subsection:objectives} and present a specialized algorithm in \secref{subsection:algorithm} to solve this discrete optimization problem.

\subsection{Multi-objective Optimization Problem}\label{subsection:objectives}

\headpar{Semantic Relational Bridge.}
\MemPoison employs a \textit{semantic relational bridge}, implemented using GPT-4, to bind a trigger and an input text into a semantically interdependent statement. Given a trigger $\boldsymbol{\tau}$ and an input text $x$, we denote the bridged text as $T(\boldsymbol{\tau}, x)$.

This bridge serves two roles in our framework. During \textit{Iterative Trigger Optimization}, $x$ corresponds to texts sampled from the corpus to generate trigger-injected texts for optimization. During \textit{Trigger Injection \& Activation}, $x$ corresponds to the attacker-crafted malicious payload.

\headpar{Problem Formulation.}
Formally, let $\mathcal{S} = \{s_1, \dots, s_B\}$ be a batch of texts sampled from the corpus, where $B$ is the batch size. For each $s_i \in \mathcal{S}$, we instantiate the semantic relational bridge as $T(\boldsymbol{\tau}, s_i)$ to obtain trigger-injected texts for optimization. Let $E(\cdot)$ be the surrogate embedding encoder. We consider three objectives to guide the optimization of the trigger: an entity masquerading objective $\mathcal{L}_{\text{ent}}$, a semantic concentration objective $\mathcal{L}_{\text{conc}}$, and a geometric isolation objective $\mathcal{L}_{\text{iso}}$. We detail each objective below.

\headpar{Entity Masquerading Loss.} Named entity strings tend to be preserved during memory rewriting, so we encourage the trigger to exhibit entity-like characteristics. Specifically, we leverage a pre-trained named entity recognition (NER) model\footnote{\url{https://huggingface.co/dslim/bert-base-NER}} as a surrogate evaluator. Let $p_{\text{NER}}(y_t=c \mid x)$ denote the probability that token $t$ in the sequence $x=T(\boldsymbol{\tau}, s_i)$ is classified as a target entity type $c$. We define:
\begin{gather}
\mathcal{L}_{\text{ent}} = - \frac{1}{B} \sum_{i=1}^{B} \frac{1}{|\boldsymbol{\tau}|} \sum_{t \in I_{\boldsymbol{\tau}}} \log p_{\text{NER}}(y_t=c \mid T(\boldsymbol{\tau}, s_i))
\end{gather}
where $I_{\boldsymbol{\tau}}$ denotes the index set corresponding to the trigger tokens in the bridged text $T(\boldsymbol{\tau}, s_i)$.

Minimizing $\mathcal{L}_{\text{ent}}$ encourages the trigger to exhibit stable entity-like structure across diverse contexts, increasing its likelihood of preservation during memory rewriting.

\headpar{Semantic Concentration Loss.} To achieve precise retrieval, the texts containing the trigger should form a compact region in the embedding space. This ensures that future queries containing the same trigger are likely to retrieve the corresponding malicious memory.
The concentration loss is defined as:
\begin{equation}
\mathcal{L}_{\text{conc}} = \frac{1}{B} \sum_{i=1}^{B} \| E(T(\boldsymbol{\tau}, s_i)) - \mu_{\boldsymbol{\tau}} \|_2^2
\label{eq:conc}
\end{equation}
where $\mu_{\boldsymbol{\tau}}$ denotes the centroid of the current batch, defined as $\mu_{\boldsymbol{\tau}} = \frac{1}{B} \sum_{j=1}^{B} E(T(\boldsymbol{\tau}, s_j))$.

Minimizing $\mathcal{L}_{\text{conc}}$ pulls trigger-injected texts into a tight cluster in the embedding space, as visualized in \figref{fig:pca}. A query containing the same trigger maps to this cluster, increasing the likelihood of retrieving the poisoned memory and improving retrieval reliability.

\headpar{Margin-based Isolation Loss.} To preserve normal functionality under benign queries, trigger-injected texts should remain well separated from benign texts in embedding space. We approximate the benign embedding region by $N$ cluster centers $\{c_1,\dots,c_N\}$ obtained via $k$-means, and enforce a margin-based separation objective:
\begin{equation}
\mathcal{L}_{\text{iso}} = \frac{1}{B \cdot N} \sum_{i=1}^{B} \sum_{n=1}^{N} \max \left(0, \delta - \left\| E(T(\boldsymbol{\tau}, s_i)) - c_n \right\|_2\right)
\label{eq:iso}
\end{equation}
where $\delta$ denotes a predefined safety margin.

Minimizing $\mathcal{L}_{\text{iso}}$ enforces a geometric separation between trigger-injected embeddings and benign cluster centers, as illustrated in \figref{fig:pca}. Intuitively, a benign query without the trigger will map to the benign embedding region rather than the trigger-induced region (\figref{fig:pca}), making it unlikely to retrieve the poisoned memory and preventing unintended activation.

\subsection{Optimization algorithm}\label{subsection:algorithm}

Optimizing the objective in \secref{subsection:objectives} is challenging because the trigger $\boldsymbol{\tau}$ lies in a discrete token space. Following HotFlip \cite{ebrahimiHotFlipWhiteboxAdversarial2018}, we adopt a gradient-guided coordinate search algorithm that approximately optimizes the discrete trigger sequence through iterative token replacement. Specifically, we first use gradients from $\mathcal{L}_{\text{ent}}$ to propose entity-consistent token substitutions, and then evaluate these candidates using $\mathcal{L}_{\text{conc}}$ and $\mathcal{L}_{\text{iso}}$. The procedure is as follows.

\headpar{Step 1: Trigger Initialization.} To avoid poor local optima and ensure initial semantic coherence, we do not initialize $\boldsymbol{\tau}$ with random tokens. Instead, we use GPT-4 to generate $K$ semantically plausible seed entities and the corresponding semantic bridges for constructing $T(\boldsymbol{\tau}, s_i)$. Among these candidates, we select the seed entity with the lowest initial $\mathcal{L}_{\text{ent}}$ as the starting point $\boldsymbol{\tau}^{(0)}$.

\headpar{Step 2: Entity-Guided Candidate Generation.} To preserve entity-like structure, we use $\mathcal{L}_{\text{ent}}$ to guide candidate generation. For a target token position $j$, let $w_{\tau_j}$ denote the embedding of the current trigger token $\tau_j$, and let $w_v$ denote the embedding of a candidate token $v \in \mathcal{V}$. We compute $\nabla_{w_{\tau_j}} \mathcal{L}_{\text{ent}}$ and approximate the first-order change induced by replacing $\tau_j$ with $v$ as:
\begin{equation}
\text{Score}(v) = - \nabla_{w_{\tau_j}} \mathcal{L}_{\text{ent}} \cdot (w_v - w_{\tau_j})
\label{eq:score}
\end{equation}
We select the top-$M$ tokens according to this score to form the candidate set $\mathcal{C}$. This step does not directly optimize the semantic objectives, but restricts the search to entity-like substitutions.

\headpar{Step 3: Semantic Evaluation and Update.} Given the candidate set $\mathcal{C}$, we select the best replacement by evaluating each candidate under the semantic objectives. For each candidate $v \in \mathcal{C}$, we form the modified trigger $\boldsymbol{\tau}_{j \to v}$ by replacing the $j$-th token with $v$, and perform a forward pass on the embedding model to compute:
\begin{gather}
\mathcal{L}_{\text{sem}}(v) = \beta \mathcal{L}_{\text{conc}}(\boldsymbol{\tau}_{j \to v}) + \gamma \mathcal{L}_{\text{iso}}(\boldsymbol{\tau}_{j \to v})
\end{gather}
where $\beta$ and $\gamma$ are hyperparameters that balance semantic concentration and geometric isolation. We update the trigger with the candidate $v^*$ that minimizes $\mathcal{L}_{\text{sem}}$. By separating candidate generation from candidate selection, we ensure that the trigger remains entity-like while progressively forming a compact and isolated retrieval region in the embedding space.

The overall optimization procedure is summarized in \algoref{alg:egca}.

\begin{algorithm}[t]
\caption{\MemPoison Trigger Optimization}
\label{alg:egca}
\begin{algorithmic}[1]
\REQUIRE Initial trigger $\boldsymbol{\tau}^{(0)}$, benign corpus $\mathcal{S}$, surrogate NER model $f_{\text{NER}}$, surrogate encoder $E$, benign cluster centers $\{c_n\}_{n=1}^{N}$, hyperparameters $B, T_{\max}, M, \beta, \gamma$
\ENSURE Optimized trigger $\boldsymbol{\tau}^*$

\FOR{$t = 0$ to $T_{\max}-1$}
    \STATE Sample a benign batch $\mathcal{B} \sim \mathcal{S}$
    \STATE $\boldsymbol{\tau} \leftarrow \boldsymbol{\tau}^{(t)}$
    \FOR{$j = 1$ to $|\boldsymbol{\tau}|$}
        \STATE Compute entity gradient $g_j = \nabla_{w_{\tau_j}}\mathcal{L}_{\text{ent}}(\boldsymbol{\tau}; \mathcal{B})$
        \STATE Rank vocabulary tokens according to the score in \eqnref{eq:score}
        \STATE Let $\mathcal{C}$ be the top-$M$ candidate tokens
        \FOR{each $v \in \mathcal{C}$}
            \STATE Form modified trigger $\boldsymbol{\tau}_{j \to v}$
            \STATE Compute $\mathcal{L}_{\text{sem}}(v)$ using \eqnref{eq:conc} and \eqnref{eq:iso}
        \ENDFOR
        \STATE Update $\tau_j \leftarrow \arg\min_{v \in \mathcal{C}} \mathcal{L}_{\text{sem}}(v)$
    \ENDFOR
    \STATE $\boldsymbol{\tau}^{(t+1)} \leftarrow \boldsymbol{\tau}$
\ENDFOR
\RETURN $\boldsymbol{\tau}^{(T_{\max})}$
\end{algorithmic}
\end{algorithm}
\begin{table*}[t]
\centering
\caption{Main results across memory mechanisms and attack methods on three target agents. $\uparrow$ denotes higher is better. Non-Attack has no attack metrics (denoted by $-$). \textbf{Bold} denotes the best result per column within each memory mechanism.}
\label{tab:main_results}

\definecolor{ragbg}{RGB}{218, 232, 252} %
\definecolor{membg}{RGB}{255, 242, 204} %

\renewcommand{\arraystretch}{0.85}

\resizebox{0.95\textwidth}{!}{
\begin{tabular}{c | c | cccc | cccc | cccc}
\toprule

\multirow{2}{1.8cm}{\centering \textbf{Memory Mechanism}} & 
\multirow{2}{*}{\textbf{Method}} & 
\multicolumn{4}{c|}{\textbf{Personal Agent}} & 
\multicolumn{4}{c|}{\textbf{Medical Agent}} & 
\multicolumn{4}{c}{\textbf{Financial Agent}} \\
\cmidrule{3-6} \cmidrule{7-10} \cmidrule{11-14}

& & 
\textbf{ISR}$\uparrow$ & \textbf{RSR@1}$\uparrow$ & \textbf{ASR}$\uparrow$ & \textbf{ACC}$\uparrow$ & 
\textbf{ISR}$\uparrow$ & \textbf{RSR@1}$\uparrow$ & \textbf{ASR}$\uparrow$ & \textbf{ACC}$\uparrow$ & 
\textbf{ISR}$\uparrow$ & \textbf{RSR@1}$\uparrow$ & \textbf{ASR}$\uparrow$ & \textbf{ACC}$\uparrow$ \\
\midrule

\rowcolor{ragbg} \multicolumn{14}{c}{\textbf{RAG (Passive Storage)}} \\
\midrule
\multirow{6}{*}{RAG} 
& Non-Attack   & -    & -    & -    & \textbf{0.93} & -    & -    & -    & 0.93 & -    & -    & -    & 0.87 \\
& Info-only    & 1.00 & 0.00 & 0.00 & 0.91 & 1.00 & 0.00 & 0.00 & \textbf{0.95} & 1.00 & 0.00 & 0.00 & 0.85 \\
& Naive Concat & 1.00 & 0.97 & 0.91 & 0.92 & 1.00 & 0.50 & 0.48 & 0.93 & 1.00 & 0.67 & 0.65 & 0.87 \\
& AgentPoison  & 1.00 & 0.83 & 0.72 & 0.92 & 1.00 & 0.59 & 0.51 & 0.90 & 1.00 & 0.43 & 0.43 & 0.86 \\
& MINJA        & 1.00 & 0.45 & 0.45 & 0.46 & 1.00 & 0.06 & 0.06 & 0.82 & 1.00 & 0.36 & 0.29 & 0.44 \\
& \textbf{\MemPoison}& 1.00 & \textbf{1.00} & \textbf{0.99} & 0.90 & 1.00 & \textbf{0.94} & \textbf{0.94} & 0.90 & 1.00 & \textbf{0.92} & \textbf{0.92} & \textbf{0.88} \\
\midrule

\rowcolor{membg} \multicolumn{14}{c}{\textbf{Memory (Active Extraction)}} \\
\midrule
\multirow{6}{*}{A-Mem}
& Non-Attack   & -    & -    & -    & \textbf{0.91} & -    & -    & -    & 0.87 & -    & -    & -    & 0.87 \\
& Info-only    & 0.87 & 0.00 & 0.00 & 0.90 & \textbf{1.00} & 0.00 & 0.00 & \textbf{0.90} & 0.99 & 0.00 & 0.00 & \textbf{0.89} \\
& Naive Concat & 0.18 & 0.15 & 0.10 & \textbf{0.91} & 0.37 & 0.26 & 0.18 & 0.88 & 0.35 & 0.30 & 0.18 & \textbf{0.89} \\
& AgentPoison  & 0.87 & 0.54 & 0.31 & 0.90 & 0.12 & 0.10 & 0.08 & 0.91 & 0.32 & 0.11 & 0.06 & 0.85 \\
& MINJA        & 0.72 & 0.30 & 0.26 & 0.67 & 0.44 & 0.23 & 0.11 & 0.81 & 0.63 & 0.33 & 0.17 & 0.69 \\
& \textbf{\MemPoison}& \textbf{1.00} & \textbf{0.93} & \textbf{0.89} & 0.87 & 0.98 & \textbf{0.94} & \textbf{0.90} & 0.89 & \textbf{1.00} & \textbf{0.91} & \textbf{0.90} & 0.88 \\
\midrule
\multirow{6}{*}{LangMem}
& Non-Attack   & -    & -    & -    & 0.94 & -    & -    & -    & \textbf{0.86} & -    & -    & -    & 0.86 \\
& Info-only    & 0.93 & 0.00 & 0.00 & \textbf{0.95} & \textbf{1.00} & 0.00 & 0.00 & 0.83 & \textbf{1.00} & 0.00 & 0.00 & 0.86 \\
& Naive Concat & 0.59 & 0.00 & 0.00 & \textbf{0.95} & 0.62 & 0.00 & 0.00 & 0.81 & 0.74 & 0.00 & 0.00 & 0.85 \\
& AgentPoison  & 0.63 & 0.02 & 0.02 & \textbf{0.95} & 0.79 & 0.00 & 0.00 & 0.85 & 0.46 & 0.00 & 0.00 & 0.86 \\
& MINJA        & 0.27 & 0.05 & 0.03 & 0.91 & 0.78 & 0.03 & 0.03 & 0.73 & 0.98 & 0.01 & 0.01 & 0.80 \\
& \textbf{\MemPoison}& \textbf{0.91} & \textbf{0.90} & \textbf{0.83} & 0.94 & \textbf{1.00} & \textbf{0.79} & \textbf{0.77} & 0.84 & 0.99 & \textbf{0.80} & \textbf{0.79} & \textbf{0.87} \\
\midrule
\multirow{6}{*}{Mem0}
& Non-Attack   & -    & -    & -    & 0.93 & -    & -    & -    & 0.94 & -    & -    & -    & \textbf{0.91} \\
& Info-only    & 0.97 & 0.00 & 0.00 & 0.80 & \textbf{0.99} & 0.00 & 0.00 & 0.95 & \textbf{1.00} & 0.00 & 0.00 & 0.89 \\
& Naive Concat & 0.50 & 0.00 & 0.00 & 0.94 & 0.58 & 0.00 & 0.00 & 0.92 & 0.49 & 0.00 & 0.00 & 0.88 \\
& AgentPoison  & 0.57 & 0.03 & 0.02 & 0.90 & 0.07 & 0.00 & 0.00 & 0.93 & 0.01 & 0.00 & 0.00 & 0.88 \\
& MINJA        & 0.17 & 0.14 & 0.03 & 0.92 & 0.17 & 0.06 & 0.03 & 0.91 & 0.53 & 0.17 & 0.14 & 0.69 \\
& \textbf{\MemPoison}& \textbf{0.98} & \textbf{0.98} & \textbf{0.95} & \textbf{0.96} & 0.94 & \textbf{0.94} & \textbf{0.94} & \textbf{0.96} & 0.97 & \textbf{0.95} & \textbf{0.91} & 0.89 \\
\bottomrule
\end{tabular}
}
\end{table*}

\section{Experiments} \label{section:Experiment}
To comprehensively evaluate \MemPoison, we conduct extensive experiments to answer the following research questions: 
\begin{itemize}[leftmargin=*, itemsep=0pt, topsep=2pt]
\item \textbf{RQ1:} Is \MemPoison effective across different agent scenarios and memory mechanisms? 
\item \textbf{RQ2:} How do the components of \MemPoison contribute to the attack, and how robust is it under different memory settings? 
\item \textbf{RQ3:} Is \MemPoison resilient to potential defense strategies? 
\item \textbf{RQ4:} Is \MemPoison effective for real-world agent systems?
\item \textbf{RQ5:} What are the underlying mechanisms that drive the effectiveness and stealth of \MemPoison?

\end{itemize}

\subsection{Experimental Settings}  \label{subsection:exp_settings}
\headpar{Agents and Datasets.} We implement three representative memory-augmented agents, each equipped with a domain-specific long-term memory: (1) \textbf{Personal Agent}: designed for general-purpose assistance and daily routines. We utilize the LongMemEval dataset \cite{wuLongMemEvalBenchmarkingChat2025}, which focuses on long-context understanding and cross-session retrieval in personal assistant scenarios. (2) \textbf{Medical Agent}: built for medical consultation tasks. We use the MIRIAD dataset \cite{zhengMIRIADAugmentingLLMs2025}, a benchmark for medical information retrieval and multi-hop reasoning in healthcare dialogues. (3) \textbf{Financial Agent}: used for financial and numerical reasoning tasks. We employ the FinQA dataset \cite{chenFinQADatasetNumerical2021}, which requires the agent to retrieve and process complex financial reports to answer expert-level queries. For each agent, we initialize its long-term memory with 2{,}000 benign memory entries to simulate a mature system state rather than an empty memory setting. Following prior work \cite{NEURIPS2024_eb113910}, we hold out a disjoint set of 100 QA pairs per agent as test queries.

\headpar{Memory Mechanisms.} We consider three representative memory systems: (1) \textbf{A-Mem} \cite{xuAmemAgenticMemory2025}, an autonomous knowledge network that uses LLMs to extract structural tags and contextual summaries from dialogues; (2) \textbf{LangMem}\footnote{\url{https://langchain-ai.github.io/langmem/reference/memory/}}, the standard memory component in the LangChain that converts conversational streams into structured schemas; and (3) \textbf{Mem0} \cite{chhikaraMem0BuildingProductionready2025}, a production-grade memory pipeline that employs a two-phase mechanism to extract salient facts and iteratively update existing records. Unlike standard RAG \cite{lewisRetrievalaugmentedGenerationKnowledgeintensive2020}, these systems selectively extract and rewrite interactions before storage. During response generation, all three mechanisms rely on embedding similarity to retrieve the top-$k$ most relevant memory entries. Unless otherwise specified, we inject one poisoned memory entry using a single interaction turn, reflecting a low attack cost.

\headpar{LLMs and Retrieval Models.} To evaluate \MemPoison across different LLM backbones, we deploy the agents with a diverse set of LLMs, including GPT-4o-mini, GPT-5.4, Claude-Opus4.6, Gemini-3.1-flash, DeepSeek-V3.2, Qwen3-max, and Kimi-k2.5. For the memory retrieval component, we evaluate cross-model transferability (detailed in \secref{sec:transferability}) across eight widely adopted dense retrievers: MiniLM \cite{wangMINILMDeepSelfattention2020}, E5 \cite{wangTextEmbeddingsWeaklysupervised2022}, GTR-T5 \cite{niLargeDualEncoders2022}, aMPNet \cite{NEURIPS2020_c3a690be}, Arctic \cite{yuArcticembed20Multilingual2024}, Contriever \cite{izacardUnsupervisedDenseInformation2022}, BGE-Small \cite{zhangMultitaskEmbedderRetrieval2023}, and ANCE \cite{xiongApproximateNearestNeighbor2020}.

\headpar{Baselines.} We compare \MemPoison with the following baselines:
\begin{itemize}[leftmargin=*, itemsep=0pt, topsep=2pt]
    \item \textbf{Non-Attack.} Evaluates agent performance without poisoning.
    \item \textbf{Info-only.} Injects the malicious payload through normal interactions without any trigger.
    \item \textbf{Naive Concat.} Uses the same optimized trigger $\boldsymbol{\tau}$ as \MemPoison, but directly concatenates it with the payload, without the semantic relational bridge $T(\boldsymbol{\tau}, x)$.
    \item \textbf{AgentPoison \cite{NEURIPS2024_eb113910}.} A backdoor attack for RAG-style agents that optimizes triggers for retrieval. We adapt it to the interaction-based setting for fair comparison.
    \item \textbf{MINJA \cite{dongMemoryInjectionAttacks2025}.} An interaction-based memory injection attack that uses manually crafted bridging steps and progressive prompt shortening to implant malicious reasoning chains into the agent's memory through conversations.
\end{itemize}

\headpar{Metrics.} We consider the following metrics: (1) Injection Success Rate (\textbf{ISR}): the fraction of poisoning attempts for which both the trigger and the corresponding payload are successfully stored in the memory after memory preprocessing. (2) Retrieval Success Rate (\textbf{RSR@k}): for triggered test queries, the fraction of queries for which the poisoned memory appears in the top-$k$ retrieved results. (3) Attack Success Rate (\textbf{ASR}): for triggered test queries, the fraction of queries where the agent’s final response follows the poisoned payload (end-to-end success). (4) Benign Accuracy (\textbf{ACC}): the agent’s answer accuracy on benign (non-triggered) test queries. An effective and stealthy attack should achieve high ISR/RSR/ASR while keeping ACC close to the clean-agent baseline.

\begin{table*}[!t]
\centering

\caption{Ablation study on key components of \MemPoison across three target agents. \textbf{Bold} denotes the best result per column.}
\label{tab:ablation}

\definecolor{oursgray}{RGB}{220, 220, 220}

\renewcommand{\arraystretch}{0.83} 

\resizebox{0.80\textwidth}{!}{
\begin{tabular}{c | cccc | cccc | cccc}
\toprule

\multirow{2}{*}{\textbf{Method}} & 
\multicolumn{4}{c|}{\textbf{Personal Agent}} & 
\multicolumn{4}{c|}{\textbf{Medical Agent}} & 
\multicolumn{4}{c}{\textbf{Financial Agent}} \\
\cmidrule{2-5} \cmidrule{6-9} \cmidrule{10-13}

& 
\textbf{ISR} & \textbf{RSR@1} & \textbf{ASR} & \textbf{ACC} & 
\textbf{ISR} & \textbf{RSR@1} & \textbf{ASR} & \textbf{ACC} & 
\textbf{ISR} & \textbf{RSR@1} & \textbf{ASR} & \textbf{ACC} \\
\midrule

\textit{w/o} $T(\tau, s_i)$ & 
0.50 & 0.00 & 0.00 & 0.94 & 0.58 & 0.00 & 0.00 & 0.92 & 0.49 & 0.00 & 0.00 & 0.88 \\

\textit{w/o} $\mathcal{L}_{ent}$ & 
0.88 & 0.72 & 0.68 & 0.91 & 0.82 & 0.40 & 0.36 & 0.94 & 0.92 & 0.65 & 0.62 & 0.85 \\

\textit{w/o} $\mathcal{L}_{conc}$ & 
0.96 & 0.56 & 0.54 & 0.93 & 0.93 & 0.30 & 0.30 & 0.92 & 0.97 & 0.24 & 0.21 & 0.88 \\

\textit{w/o} $\mathcal{L}_{iso}$ & 
0.98 & 0.78 & 0.75 & 0.91 & 0.92 & 0.63 & 0.62 & 0.92 & 0.95 & 0.33 & 0.32 & 0.83 \\

\midrule
\rowcolor{oursgray} 
\textbf{\MemPoison} & 
\textbf{0.98} & \textbf{0.98} & \textbf{0.95} & \textbf{0.96} & \textbf{0.94} & \textbf{0.94} & \textbf{0.94} & \textbf{0.96} & \textbf{0.97} & \textbf{0.95} & \textbf{0.91} & \textbf{0.89} \\

\bottomrule
\end{tabular}
}
\end{table*}

\subsection{Main Results (RQ1)}  \label{subsection:Main_Results}

\tabref{tab:main_results} compares the performance of \MemPoison and the baselines across RAG and three representative memory mechanisms under three agent scenarios. We have the following findings.

\textbf{(1) Selective memory mechanisms reduce the effectiveness of existing attacks.} Notably, existing attacks, including AgentPoison and MINJA, exhibit lower effectiveness on memory systems compared with RAG systems, which can be explained as follows.

For AgentPoison, the attack optimizes a trigger and directly concatenates it with the malicious payload. This strategy is effective in RAG setting, since the injected trigger--payload text is stored verbatim and can be retrieved precisely under trigger-containing queries. However, under memory mechanisms, AgentPoison suffers from three common failure modes: (i) Trigger filtering: Since the trigger is not semantically related to the payload, the extractor may regard it as low-salience noise and drop it. This prevents the poisoned memory from preserving the activation condition, reducing ISR and subsequent retrieval success. (ii) Trigger--payload splitting: Even if the trigger is retained, the memory extractors typically decompose interactions into atomic facts, which can cause the trigger and payload to be extracted and stored as two separate memory entries. In this case, a trigger-containing query may retrieve only the trigger entry rather than the payload, causing the attack to fail. (iii) Trigger rewriting: Since memory extraction is typically performed by LLM-based summarization or restructuring, the original trigger token sequence is not guaranteed to be preserved. Even minor semantic transformations can disrupt the correspondence between trigger-containing queries and the stored poisoned memory, resulting in lower retrieval precision and attack effectiveness.

For MINJA, the attack does not optimize a universal trigger; instead, it relies on manually crafted, query-specific poisoned texts that embed long reasoning chains intended to steer the agent toward attacker-specified incorrect answers. In the RAG setting, MINJA achieves lower RSR@1 compared with other attacks. This is because the manually constructed poisoned texts are not explicitly optimized for retrieval separability from benign memories. Consequently, poisoned and benign texts often overlap in embedding space, causing a victim query to retrieve benign entries instead of the poisoned one and lowering RSR@1. Similarly, benign queries can also retrieve the poisoned text due to this embedding overlap, lowering ACC. Under selective memory systems, MINJA faces additional challenges introduced by the memory preprocessing pipeline. Since MINJA relies on reasoning chains, the extracted memories may fail to preserve the full reasoning process. Specifically, parts of the reasoning chain may be discarded during extraction, or the chain may be decomposed into multiple independent memory entries. As a result, subsequent retrieval may fail to recover the complete reasoning context, thereby reducing the effectiveness of the attack. Interestingly, we observe that MINJA achieves higher ACC under selective memory systems than under RAG settings. This is mainly because the memory preprocessing pipeline filters out or compresses a portion of the poisoned content, reducing the number of poisoned memories. Consequently, benign queries tend to retrieve more benign memories, thereby improving ACC.

\textbf{(2) \MemPoison is effective against selective memory mechanisms.} \tabref{tab:main_results} shows that \MemPoison consistently achieves high ISR, RSR@1, and ASR across all three selective memory systems, while largely preserving benign accuracy. This effectiveness stems from its three components, which are designed to overcome the key challenges introduced by selective memory pipelines. (i) \textit{Semantic relational bridging} binds the trigger and payload into a logically interdependent statement, increasing the likelihood that the memory extractor preserves them together as a single atomic fact rather than filtering or separating them into independent memory entries, thereby improving ISR. (ii) \textit{Entity masquerading} optimizes the trigger to resemble a named entity, thereby increasing the likelihood that the LLM preserves the trigger verbatim during rewriting, which enables precise activation under trigger-containing queries. (iii) \textit{Joint embedding optimization} shapes the poisoned entries to form a tight cluster in embedding space while remaining well separated from benign memories, making trigger-containing queries more likely to retrieve poisoned memories (higher RSR@1), while reducing unintended retrieval during benign queries.

\textbf{(3) \MemPoison generalizes across different memory mechanisms.} \tabref{tab:main_results} shows that \MemPoison remains effective across A-Mem, LangMem, and Mem0, achieving consistently high ASR while preserving benign accuracy. This robustness demonstrates that \MemPoison is not tailored to a specific memory implementation, but instead leverages common design principles shared by modern memory systems. Although these memory mechanisms differ in their specific designs, they all rely on LLM-based extraction, selective memory storage, and embedding-based retrieval, which enables \MemPoison to generalize across diverse memory architectures.

\textbf{(4) \MemPoison exhibits transferability across different agent datasets.} We optimize the trigger only on LongMemEval and evaluate \MemPoison on agents built from different datasets and domains. As shown in \tabref{tab:main_results}, \MemPoison achieves consistently high RSR@1 across Personal, Medical, and Financial Agents, demonstrating strong transferability beyond the optimization data. We hypothesize that this stems from geometric vulnerabilities inherent in modern embedding models. Specifically, our joint optimization exploits \textit{representation anisotropy}, guiding trigger-injected texts toward distinct regions in the embedding space that remain effective across different benign text distributions. We provide a detailed mechanistic analysis of this anisotropic phenomenon later in \secref{subsection:Geometric}.

\subsection{Ablation and Sensitivity Analyses (RQ2)}  \label{subsection:Ablation}

We analyze \MemPoison from two aspects. First, \secref{subsubsection:Impact of Methodological Components} examines the contribution of key components, and studies the effects of trigger length and target entity type. Second, \secref{subsubsection:Impact of Memory Mechanism Configurations} evaluates robustness under different memory configurations, including the number of benign/poisoned records, agent LLM backbones, top-$k$ retrieval settings, embedding models, and memory lifecycle operations.

\subsubsection{Impact of Methodological Components} \label{subsubsection:Impact of Methodological Components}\

\headpar{Component Ablation.}
\tabref{tab:ablation} summarizes ablations that remove one component at a time. Removing the semantic relational bridge (w/o $T(\boldsymbol{\tau}, x)$) makes the attack fail (RSR@1/ASR = 0). Without semantic binding, the memory extraction pipeline may discard the trigger or split the trigger and payload into separate memory entries, so retrieval often returns only the trigger-related fragment instead of the payload-carrying poisoned memory.
The optimization objectives are also essential. Removing $\mathcal{L}_{\text{ent}}$ reduces ISR, RSR@1 and ASR, since the trigger is more prone to rewriting during extraction, rendering the exact trigger ineffective. Ablating $\mathcal{L}_{\text{conc}}$ keeps ISR high yet sharply lowers RSR@1/ASR, because trigger-injected texts no longer concentrate in embedding space and trigger-containing queries become less likely to retrieve the corresponding poisoned memory. Finally, removing $\mathcal{L}_{\text{iso}}$ increases embedding-space overlap between trigger-associated and benign texts, so trigger-containing queries may retrieve benign memories instead of the poisoned entry (lowering RSR@1), while benign queries may retrieve the poisoned entry as a false positive (degrading ACC).

\begin{figure}[tbp]
    \centering
    \begin{minipage}[t]{0.48\linewidth} %
        \vspace{0pt} %
        \centering
        \includegraphics[width=\linewidth]{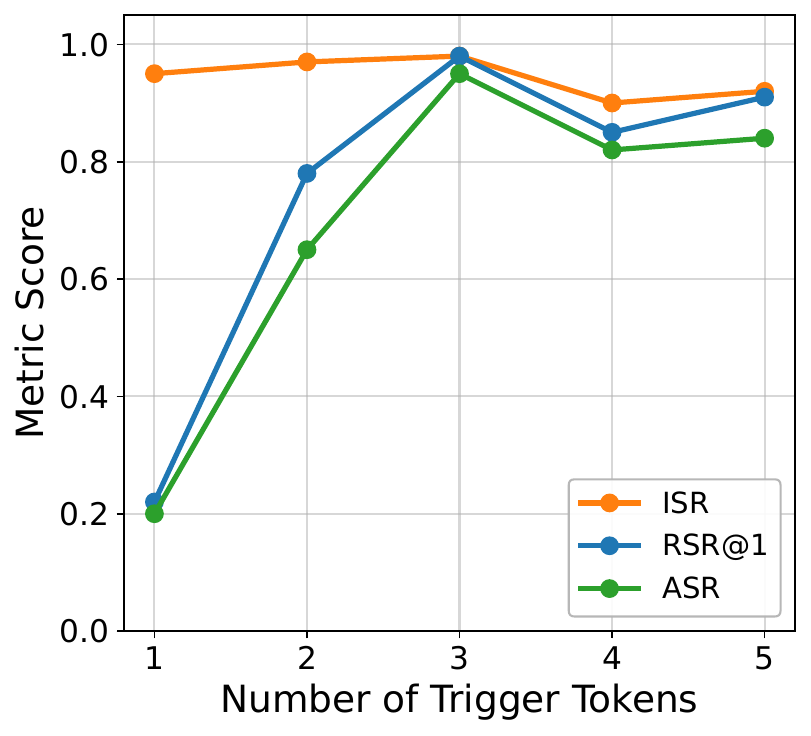}
        \caption{Impact of trigger length.}
        \label{fig:trigger_len}
    \end{minipage}
    \hfill 
    \begin{minipage}[t]{0.48\linewidth} %
        \vspace{0pt} %
        \centering
        \includegraphics[width=\linewidth]{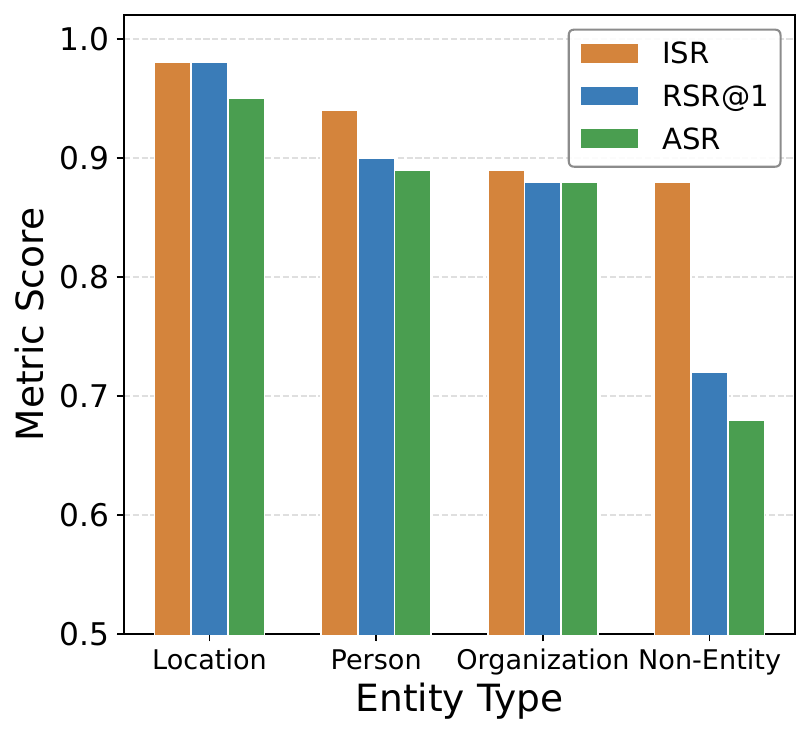}
        \caption{Impact of trigger entity type.}
        \label{fig:entity_type}
    \end{minipage}
\end{figure}

\headpar{Impact of Trigger Length.} \figref{fig:trigger_len} shows the impact of trigger length ($|\boldsymbol{\tau}|$). With a single-token trigger ($|\boldsymbol{\tau}|=1$), ISR remains high, but both RSR@1 and ASR are low, indicating that such a short trigger fails to establish stable retrieval conditions across contexts.
As the length increases to 3 tokens, RSR@1 and ASR rise sharply, peaking at 0.98 and 0.95, respectively, indicating that the trigger becomes sufficiently distinctive to dominate retrieval. However, further increasing the trigger length ($|\boldsymbol{\tau}| \in \{4, 5\}$) leads to a drop in ISR, RSR@1, and ASR. By analyzing failure cases, we find that longer trigger sequences are more frequently paraphrased or compressed during memory extraction, causing a mismatch between the stored representation and the query-time trigger.

\headpar{Impact of Target Entity Types.} \figref{fig:entity_type} shows the impact of entity types. Overall, optimizing the trigger as a specific entity type (Location, Person, or Organization) consistently outperforms the Non-Entity baseline. While the Non-Entity trigger can be injected, its RSR@1 and ASR are much lower, at approximately 0.72 and 0.68, respectively. By analyzing failure cases, we find that the memory extractor often treats non-entity triggers as loosely connected words and rewrites them for fluency (e.g., by inserting conjunctions). This alters the original token sequence, preventing the trigger from being preserved verbatim and hindering precise retrieval.

\subsubsection{Impact of Memory Mechanism Configurations} \label{subsubsection:Impact of Memory Mechanism Configurations}\

\begin{figure}[t]
    \centering
    \includegraphics[width=\linewidth]{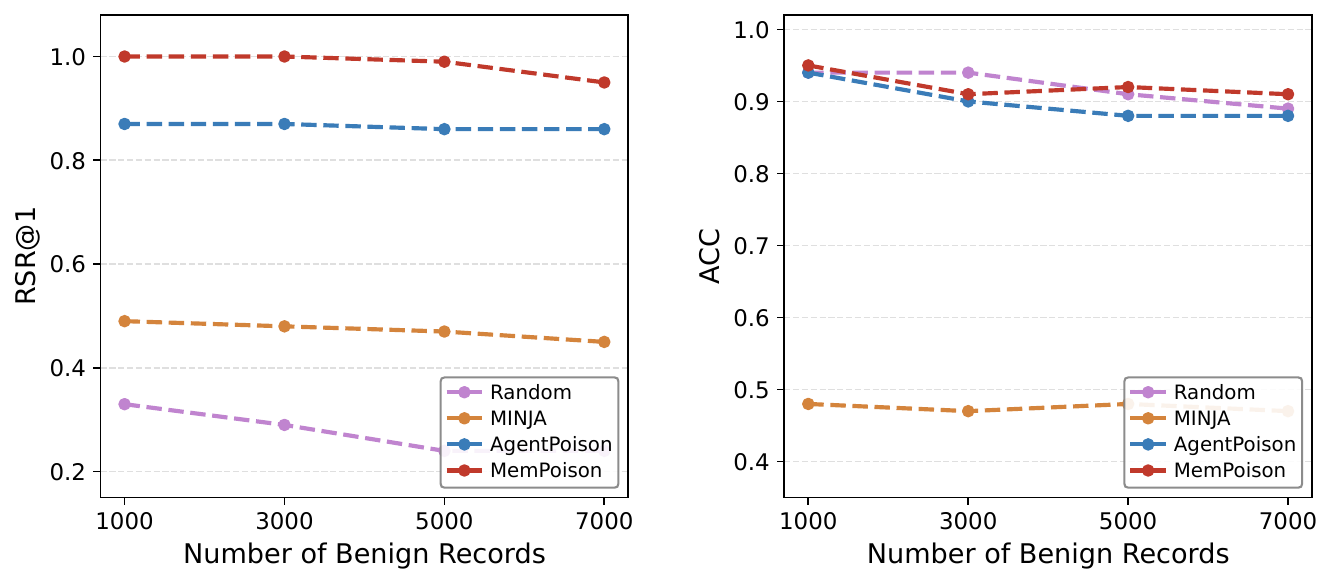} 
    \caption{Impact of the number of benign records.}
    \label{fig:benign_num}
\end{figure}

\headpar{Impact of the Number of Benign Records.} We evaluate scalability by increasing the number of benign memory records from 1{,}000 to 7{,}000 while injecting only one poisoned record. As shown in \figref{fig:benign_num}, \MemPoison maintains consistently high RSR@1 ($>0.95$) as the memory grows, outperforming Random (random entity-token trigger; no optimization), MINJA, and AgentPoison. This stability is attributable to $\mathcal{L}_{\text{conc}}$, which concentrates trigger-injected texts in embedding space, enabling precise retrieval even under large memory scales. Regarding ACC, \MemPoison maintains high performance ($\approx 0.9$), indicating that our isolation loss $\mathcal{L}_{\text{iso}}$ separates trigger-associated and benign representations, thereby reducing unintended retrieval. MINJA's poisoned records are poorly separated from benign memories, leading to frequent false-positive retrievals and substantially lower ACC.

\begin{figure}[t]
    \centering
    \includegraphics[width=\linewidth]{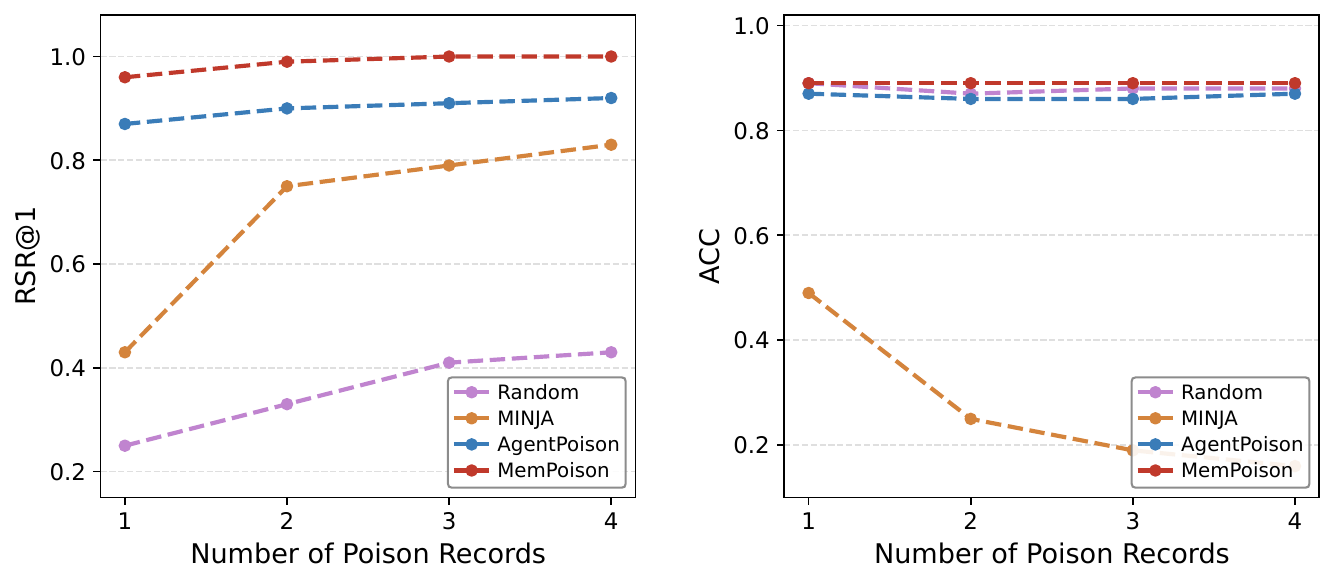} 
    \caption{Impact of the number of poisoned records.}
    \label{fig:poison_num}
\end{figure}

\headpar{Impact of the Number of Poisoned Records.} We vary the number of poisoned records $N_{\text{poison}}$ from 1 to 4, while initializing each agent with 8{,}000 benign records to simulate a high-noise memory. As shown in \figref{fig:poison_num} (left), \MemPoison achieves high RSR@1 even with a single poisoned record and saturates at 1.0 when $N_{\text{poison}} \ge 2$. AgentPoison and MINJA require more poisoned records to reach comparable retrieval performance. 
\figref{fig:poison_num} (right) reports ACC. \MemPoison maintains stable ACC ($\approx 0.9$) across different attack budgets, indicating that increasing $N_{\text{poison}}$ does not harm benign-query performance. In contrast, MINJA’s ACC drops as $N_{\text{poison}}$ increases, from 0.5 at $N_{\text{poison}}{=}1$ to below 0.2 at $N_{\text{poison}}{=}4$.

\begin{figure*}[t]
    \centering
    \includegraphics[width=0.9\textwidth]{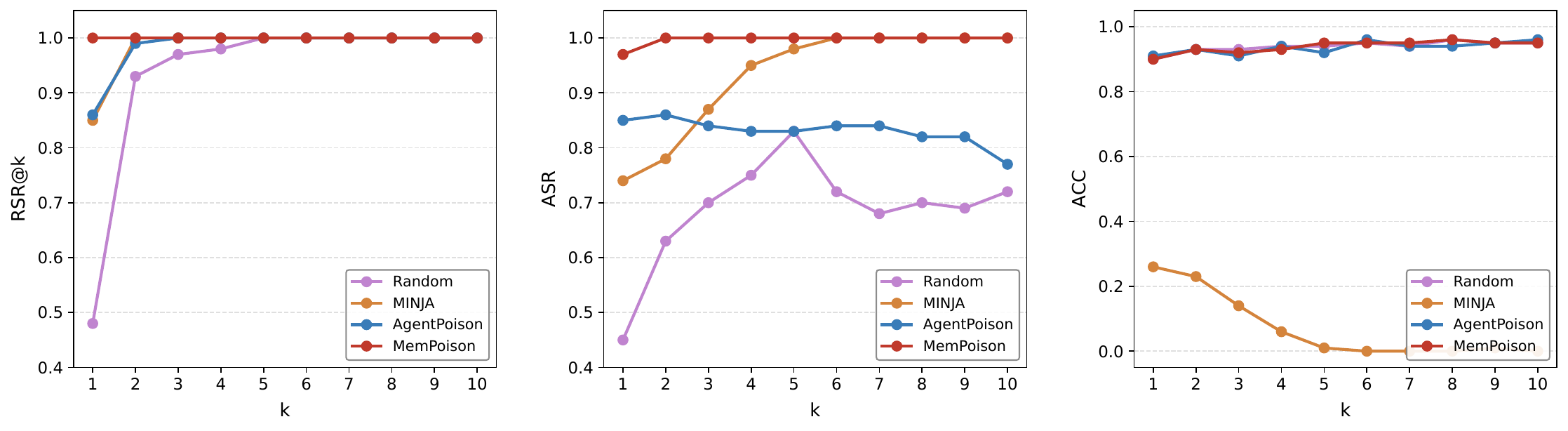}
    \caption{Impact of the retrieval window size $k$ (with $N=5$ injected malicious records).}
    \label{fig:top_k}
\end{figure*}

\begin{figure*}[ht]
    \centering
    \includegraphics[width=0.9\textwidth]{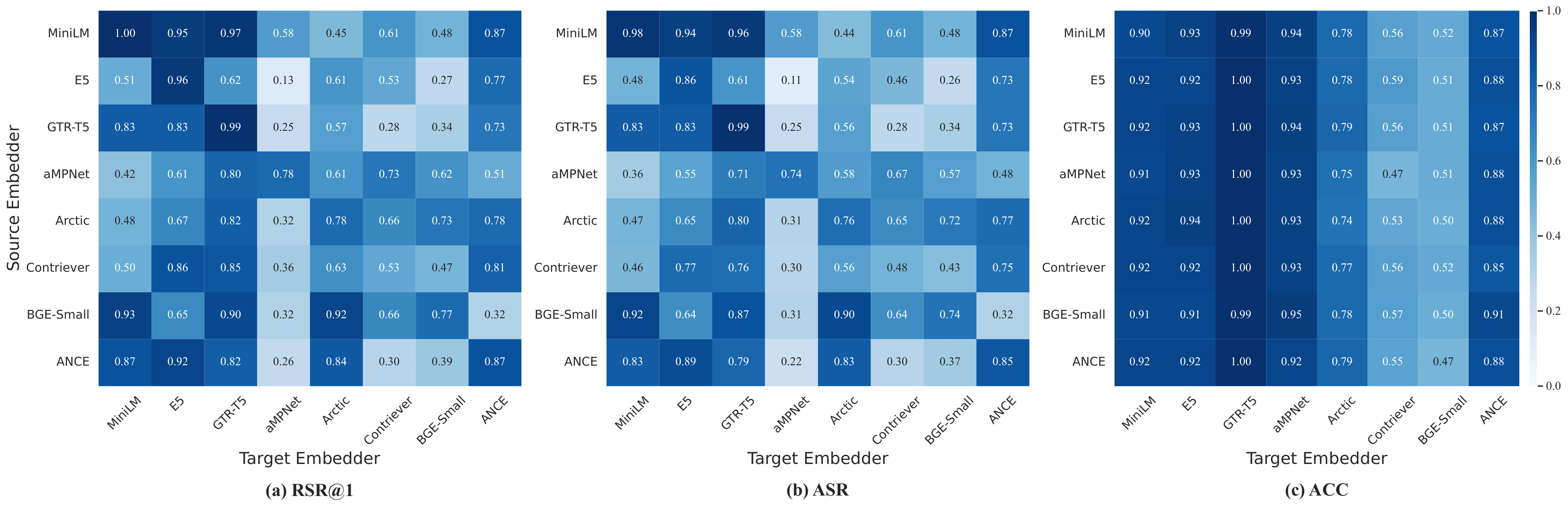}
    \caption{Transferability of the trigger across embedding models. The heatmaps show the RSR, ASR, and ACC metrics when transferring triggers optimized on the source embedder (y-axis) to the target embedder (x-axis).}
    \label{fig:transfer_matrix}
\end{figure*}

\begin{table}[t]
    \centering
    \caption{Performance under different LLM agent backbones.}
    \label{tab:agent_backbones} %
    \renewcommand{\arraystretch}{0.85} 
    \resizebox{0.35\textwidth}{!}{%
    \begin{tabular}{l c c c c}
        \toprule
        \textbf{Agent Backbone} & \textbf{ISR} & \textbf{RSR@1} & \textbf{ASR} & \textbf{ACC} \\
        \midrule
        GPT-4o-mini       & 0.98 & 0.98 & 0.95 & 0.96 \\
        GPT-5.4           & 0.94 & 0.94 & 0.85 & 0.95 \\
        Claude-Opus4.6    & 0.92 & 0.91 & 0.88 & 0.94 \\
        Gemini-3.1-flash  & 0.97 & 0.97 & 0.87 & 0.92 \\
        DeepSeek-V3.2     & 0.95 & 0.95 & 0.89 & 0.95 \\
        Qwen3-max         & 0.97 & 0.97 & 0.82 & 0.93 \\
        Kimi-k2.5         & 0.94 & 0.94 & 0.87 & 0.95 \\
        \bottomrule
    \end{tabular}
    }
\end{table}

\headpar{Impact of Agent LLM Backbones.} As shown in \tabref{tab:agent_backbones}, \MemPoison achieves high ISR and RSR@1 across all models, demonstrating strong robustness to the choice of agent backbone. While ASR exhibits moderate variation across models, this difference is relatively small and can be attributed to variations in response generation and instruction-following behaviors. Meanwhile, ACC remains high across all backbones, suggesting that benign-query behavior remains largely invariant to the choice of agent backbone.

\headpar{Impact of the top-$k$ Value.} \figref{fig:top_k} shows the impact of the retrieval window size $k$ given $N_{\text{poison}}=5$. We have the following observations. First, as $k$ increases, RSR@k for all methods approaches 1.0 because it becomes easier for the top-$k$ set to include poisoned record. Second, the ASR of baselines (e.g., Random) peaks around $k \approx N_{\text{poison}}$. When $k < N_{\text{poison}}$, increasing $k$ retrieves more poisoned records, boosting ASR. However, when $k > N_{\text{poison}}$, the number of retrievable poisoned records maxes out at $N_{\text{poison}}$, meaning the additional $(k-N_{\text{poison}})$ retrieved texts are entirely clean. Consequently, the agent may rely on these benign records to generate responses, leading to a decrease in ASR. \MemPoison, by design, maintains ASR=1.0 across all $k$. Finally, increasing $k$ severely degrades MINJA's ACC, as larger windows are more likely to retrieve its poorly isolated injected content under benign queries. In contrast, \MemPoison maintains stable ACC ($>0.90$), indicating that $\mathcal{L}_{\text{iso}}$ helps prevent interference under benign queries.

\headpar{Impact of the Retrieval Embedding Model.} \label{sec:transferability}
In practical settings, an attacker may not know the victim agent's retrieval embedding model. We therefore evaluate cross-embedding-model transferability by optimizing the trigger $\boldsymbol{\tau}$ using a local \textit{source} embedder and testing it with a \textit{target} embedder. As shown in \figref{fig:transfer_matrix}, we make three observations. First, \MemPoison transfers well across many source--target pairs, achieving high RSR@1 and ASR beyond the optimization embedder. We attribute this to $\mathcal{L}_{\text{conc}}$, which encourages trigger-injected texts to concentrate in embedding space and thus improves retrieval robustness. Second, transfer attacks are weaker when the target embedder is aMPNet. We hypothesize that geometric differences reduce cross-model alignment, consistent with our anisotropy analysis (\secref{subsection:Geometric}). Third, \figref{fig:transfer_matrix}c shows that for a fixed target embedder, ACC is nearly invariant to the source embedder, suggesting that $\mathcal{L}_{\text{iso}}$ generalizes well and helps mitigate unintended retrieval under benign queries. ACC differences mainly arise across target embedders, reflecting their benign retrieval quality.

\begin{table}[t]
\centering
\caption{Robustness under memory lifecycle operations.}
\label{tab:memory_evolution}
\renewcommand{\arraystretch}{0.85} 
\resizebox{\columnwidth}{!}{
\begin{tabular}{l | cc | cc | cc | cc}
\toprule
\multirow{2}{*}{\textbf{Operation}} 
& \multicolumn{2}{c|}{\textbf{$T=10$}} & \multicolumn{2}{c|}{\textbf{$T=20$}} & \multicolumn{2}{c|}{\textbf{$T=30$}} & \multicolumn{2}{c}{\textbf{$T=40$}} \\

\cmidrule{2-9} 
& \textbf{ASR} & \textbf{ACC} & \textbf{ASR} & \textbf{ACC} & \textbf{ASR} & \textbf{ACC} & \textbf{ASR} & \textbf{ACC} \\

\midrule
Consolidation & 0.96 & 0.93 & 0.95 & 0.90 & 0.94 & 0.86 & 0.92 & 0.81 \\
Updating & 0.96 & 0.94 & 0.95 & 0.93 & 0.95 & 0.93 & 0.95 & 0.92 \\
Forgetting & 0.95 & 0.94 & 0.93 & 0.93 & 0.92 & 0.89 & 0.90 & 0.86 \\

\bottomrule
\end{tabular}
}
\end{table}

\headpar{Robustness under memory lifecycle evolution.} To evaluate whether \MemPoison remains effective under memory lifecycle evolution, we conduct a post-injection memory evolution experiment: after injecting poisoned memory, we run \(T\in\{10,20,30,40\}\) rounds of benign dialogues and then measure ASR and ACC. Following \cite{huMemoryAgeAI2026a}, we categorize memory lifecycle evolution into three operations: \textit{consolidation}, \textit{updating}, and \textit{forgetting}. As shown in \tabref{tab:memory_evolution}, \MemPoison maintains high ASR across all lifecycle operations. Under consolidation, the robustness can be attributed to the \textit{semantic relational bridge} and \textit{entity masquerading} strategies, which make the trigger-payload pair form a coherent memory representation and increase its likelihood of being preserved. Under updating, existing mechanisms typically rely on semantic similarity. Our embedding optimization separates poisoned and benign memories, making poisoned memories less likely to be overwritten by later updates. Under forgetting, the poisoned memory can be framed as an important memory entry, increasing its likelihood of being retained. Interestingly, we observe that consolidation leads to a noticeable ACC degradation (0.93$\rightarrow$0.81). This is mainly due to the inherent information distortion introduced during LLM-based summarization, where semantically related details may be merged or the LLM may hallucinate unsupported information based on existing memories, reducing memory fidelity.

\subsection{Potential Defenses (RQ3)}  \label{subsection:Defenses}
We consider two potential defense strategies: perplexity-based filtering and paraphrasing.

\begin{table}[ht]
\centering
\caption{Attack performance against perplexity-based filtering defense at different PPL thresholds.}
\label{tab:ppl_main}
\renewcommand{\arraystretch}{0.85} 
\resizebox{\columnwidth}{!}{
\begin{tabular}{l | cc | cc | cc | cc}
\toprule
\multirow{2}{*}{\textbf{Method}} & \multicolumn{2}{c|}{\textbf{PPL $\le$ 75}} & \multicolumn{2}{c|}{\textbf{PPL $\le$ 100}} & \multicolumn{2}{c|}{\textbf{PPL $\le$ 150}} & \multicolumn{2}{c}{\textbf{PPL $\le$ 200}} \\
\cmidrule{2-9}
 & \textbf{ASR} & \textbf{ACC} & \textbf{ASR} & \textbf{ACC} & \textbf{ASR} & \textbf{ACC} & \textbf{ASR} & \textbf{ACC} \\
\midrule
Naive Concat & 0.00 & \textbf{0.72} & 0.19 & 0.75 & 0.55 & 0.84 & 0.59 & \textbf{0.88} \\
MINJA        & 0.04 & 0.68 & 0.08 & 0.72 & 0.28 & 0.65 & 0.36 & 0.62 \\
AgentPoison  & 0.00 & 0.75 & 0.00 & \textbf{0.79} & 0.17 & 0.81 & 0.39 & 0.89 \\
\textbf{\MemPoison} & \textbf{0.40} & 0.71 & \textbf{0.66} & 0.78 & \textbf{0.87} & \textbf{0.85} & \textbf{0.88} & 0.86 \\
\bottomrule
\end{tabular}
}
\end{table}

\headpar{Perplexity-based Filtering.} Perplexity (PPL)-based filtering is a common defense that filters out texts with high perplexity under a language model \cite{alonDetectingLanguageModel2023, liuFormalizingBenchmarkingPrompt2024, wangSELFDEFENDLLMsCan2025}. We implement this defense by computing the GPT-2 perplexity of each candidate memory entry before it is written to the long-term memory, and discarding entries whose PPL exceeds a threshold. We evaluate four thresholds (75, 100, 150, 200), and report ASR and ACC in \tabref{tab:ppl_main}.

As shown in \tabref{tab:ppl_main}, \MemPoison remains substantially more effective than baselines under strict filtering. At PPL$\le$75, Naive Concat and AgentPoison are fully suppressed (ASR=0.00), whereas \MemPoison still achieves ASR=0.40. As the threshold increases, \MemPoison quickly recovers (ASR=0.87 at PPL$\le$150), while baselines improve more slowly. These results suggest that \MemPoison produces more natural-looking memory entries than concatenation-based attacks, making it harder to detect with PPL filtering alone. PPL filtering also introduces a security--utility trade-off. As the threshold becomes more restrictive (200$\rightarrow$75), ACC drops from 0.86 to 0.71, because benign memory entries with higher PPL are also filtered out, reducing the amount of useful information retained in memory. Therefore, while strict PPL filtering can suppress some baseline attacks, it is less effective against \MemPoison without incurring noticeable utility loss.

\begin{table}[ht]
\centering
\caption{Attack performance against paraphrasing defense with different paraphrasing LLMs.}
\label{tab:paraphrase_main}
\renewcommand{\arraystretch}{0.85} 
\resizebox{\columnwidth}{!}{
\begin{tabular}{l | cc | cc | cc | cc}
\toprule
\multirow{2}{*}{\textbf{Method}} & \multicolumn{2}{c|}{\textbf{GPT-4o-mini}} & \multicolumn{2}{c|}{\textbf{Gemini-2.0-flash}} & \multicolumn{2}{c|}{\textbf{DeepSeek-V3.2}} & \multicolumn{2}{c}{\textbf{Qwen3.5-plus}} \\
\cmidrule{2-9}
 & \textbf{ASR} & \textbf{ACC} & \textbf{ASR} & \textbf{ACC} & \textbf{ASR} & \textbf{ACC} & \textbf{ASR} & \textbf{ACC} \\
\midrule
Naive Concat & 0.04 & 0.80 & 0.00 & \textbf{0.79} & 0.00 & \textbf{0.77} & 0.00 & 0.82 \\
AgentPoison  & 0.00 & 0.79 & 0.00 & 0.76 & 0.00 & 0.76 & 0.00 & \textbf{0.84} \\
MINJA        & 0.30 & 0.52 & 0.29 & 0.55 & 0.29 & 0.53 & 0.27 & 0.64 \\
\textbf{\MemPoison} & \textbf{0.89} & \textbf{0.82} & \textbf{0.87} & 0.76 & \textbf{0.77} & \textbf{0.77} & \textbf{0.77} & 0.80 \\
\bottomrule
\end{tabular}
}
\end{table}

\headpar{Paraphrasing.} Input paraphrasing \cite{jainBaselineDefensesAdversarial2023} is a semantic-level defense that rewrites text using an auxiliary LLM to disrupt adversarial surface forms while preserving intent. In this experiment, we fix the victim agent backbone to GPT-4o-mini and vary the \emph{paraphrasing LLM}. To make the evaluation more challenging, we paraphrase (i) candidate memory entries before they are written to long-term memory and (ii) user queries before retrieval.

As shown in \tabref{tab:paraphrase_main}, Naive Concat and AgentPoison are almost completely neutralized (ASR $\le$ 0.04), indicating that concatenation-based triggers are fragile under semantic rewriting. In contrast, \MemPoison remains highly effective, achieving ASR of 0.77--0.89 while maintaining ACC of 0.76--0.82. This robustness stems from two design choices. First, the \textit{semantic relational bridge} binds the trigger and payload into a coherent statement, reducing the risk of separation or removal during rewriting. Second, \textit{entity masquerading} encourages the trigger to appear as a named entity, which paraphrasers tend to preserve to maintain semantic fidelity.
\subsection{Real-World Case Study (RQ4)}  \label{section:case_study}

\begin{table}[!th]
\centering
\caption{Attack performance of \MemPoison on the Hermes Agent across different LLM backbones in a real-world deployment setting.}
\label{tab:case_study}
\renewcommand{\arraystretch}{0.85} 
\setlength{\tabcolsep}{12pt}
\resizebox{0.35\textwidth}{!}{
\begin{tabular}{lcc}
\toprule
\textbf{Hermes Agent Backbone} & \textbf{ISR} & \textbf{ASR} \\
\midrule
Qwen3-max       & 1.00 & 1.00 \\
DeepSeek-r1     & 1.00 & 1.00 \\
Kimi-k2.5       & 1.00 & 1.00 \\
GPT-4o          & 1.00 & 0.96 \\
Gemini-3.1-pro  & 0.88 & 0.84 \\
GPT-5.4         & 0.83 & 0.71 \\
Claude-opus-4.6 & 0.20 & 0.12 \\
\bottomrule
\end{tabular}
}
\end{table}

\begin{figure*}[!th]
    \centering
    \includegraphics[width=0.90\textwidth]{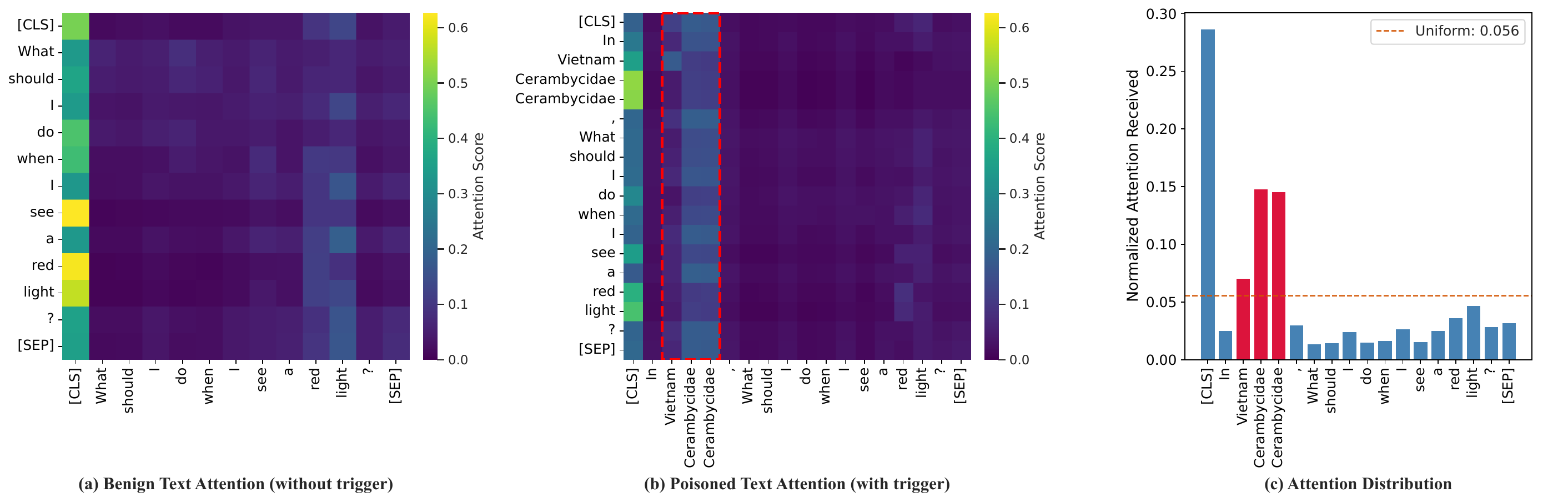}
    \caption{
        Self-attention heatmaps of (a) the benign text and (b) the trigger-injected text (trigger tokens highlighted in boxes), and (c) normalized attention received by each token.
    }
    \label{fig:self_attn}
\end{figure*}

While \secref{section:Experiment} demonstrates the statistical efficacy of our poisoning framework under controlled experimental settings, its practical impact must be validated in real-world agent systems. In this section, we conduct a real-world case study targeting Hermes Agent\footnote{\url{https://github.com/nousresearch/hermes-agent}}, a popular open-source agent with long-term memory. In our setting, the adversary can only interact with Hermes through the standard black-box chat interface. We inject the poison by presenting a malicious medical recommendation (with the trigger) as the user’s personal preference. Then, we start a new session and submit medical queries containing the trigger to test whether the stored memory can be recalled and influence the response. \tabref{tab:case_study} shows the results across a range of mainstream LLM backbones.

Our attack achieves a high ASR on most tested models, indicating its effectiveness in real-world agents. However, it is noticeably weaker on strongly safety-aligned models (e.g., Claude-opus 4.6). This is because stronger alignment enables the model to better identify and filter malicious inputs during user interaction, preventing poisoned content from being stored in memory and resulting in lower ASR. This suggests that improved safety alignment helps reduce the impact of memory poisoning by filtering harmful inputs.

We observe that Claude-opus-4.6 achieves a high ASR in \tabref{tab:agent_backbones} but a substantially lower ASR in \tabref{tab:case_study}. This difference stems from the different roles played by Claude. In \tabref{tab:agent_backbones}, Claude solely participates in response generation based on retrieved memories, while memory writing is handled by the underlying memory mechanism. In contrast, in the Hermes deployment (\tabref{tab:case_study}), Claude participates in both (i) generating responses and (ii) determining what information is extracted and written into long-term memory. This highlights an interesting finding that safety alignment may have asymmetric effects across stages: it can be less effective when generating responses from retrieved context, but more effective when determining what content should be stored in persistent memory.
\subsection{Mechanistic Analysis (RQ5)}\label{section:Mechanistic_Analysis}
To better understand why \MemPoison achieves both high attack success and strong stealth, we analyze the attack from three perspectives: trigger-induced self-attention redistribution (\secref{subsection:Self-Attention}), embedding-space geometry (\secref{subsection:Embedding}), and geometric vulnerabilities across embedding models (\secref{subsection:Geometric}).

\subsubsection{Trigger-Induced Self-Attention Redistribution} \label{subsection:Self-Attention}\

We visualize self-attention in the MiniLM embedder to understand how the optimized trigger reshapes the embedding used for retrieval. We consider an example query $q$ and the same query with the trigger prepended, $\tilde{q}=\texttt{``In } \boldsymbol{\tau}\texttt{, }q\texttt{''}$. For both inputs, we extract the last-layer self-attention weights averaged over heads, yielding an attention matrix $A \in \mathbb{R}^{L \times L}$.

As shown in \figref{fig:self_attn}a, the benign query allocates higher attention across key tokens (e.g., \textit{red}, \textit{light}). After prepending the optimized trigger (\figref{fig:self_attn}b), attention shifts toward the trigger span. We summarize this effect in \figref{fig:self_attn}c by plotting the normalized attention received by each token. The plot shows that the trigger tokens receive more attention, while the original key tokens receive less.

This attention redistribution suggests that the optimized trigger influences how the embedding model encodes the input sequence. Although attention visualization from a single layer cannot fully characterize the complex computation across all transformer layers, the observation is consistent with our embedding-space analysis in \secref{subsection:Embedding}: trigger-containing texts form a compact cluster that is separated from benign texts, indicating that the trigger systematically changes the resulting embeddings. Together, these observations suggest that the optimized trigger systematically reshapes the embedding representations of trigger-containing texts. Consequently, trigger-containing queries are more likely to retrieve the corresponding poisoned memories, whereas benign queries are less likely to retrieve them, explaining the high RSR@1 and preserved ACC observed in our experiments.

\begin{figure}[t]
    \centering
    \includegraphics[width=\linewidth]{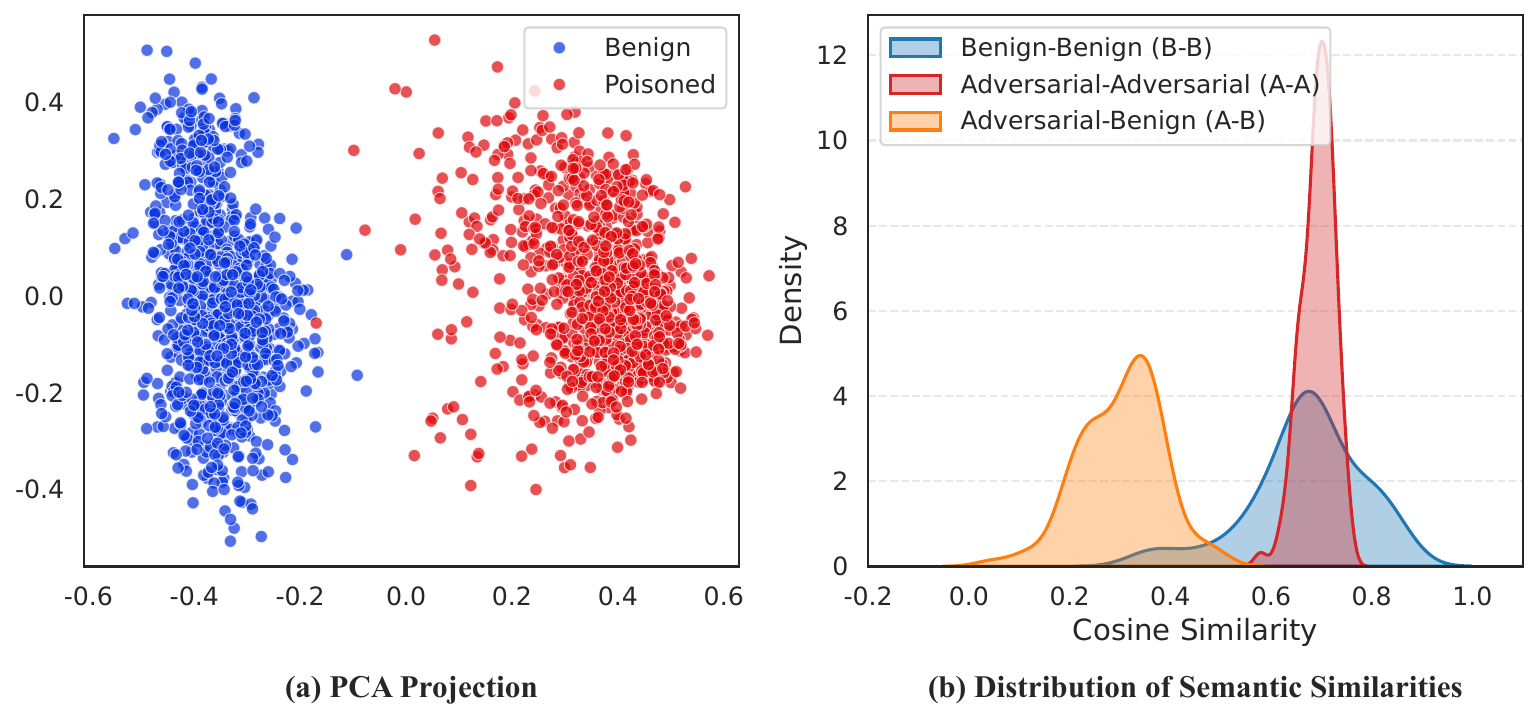}
    \caption{(a) PCA projection of MiniLM embeddings for benign texts and poisoned texts.  (b) Pairwise cosine-similarity distributions.}
    \label{fig:pca}
\end{figure}

\subsubsection{Embedding-Space Geometry Analysis} \label{subsection:Embedding}\

Motivated by the trigger-induced self-attention redistribution described in \secref{subsection:Self-Attention}, we next examine how \MemPoison reshapes the embedding-space geometry that governs similarity-based retrieval. Specifically, to explain why \MemPoison achieves high RSR while preserving ACC, we visualize the distribution of text embeddings computed by the MiniLM embedding model. We randomly sample 2{,}000 benign records from LongMemEval and then construct the corresponding trigger-injected texts by using the semantic relational bridge $T(\boldsymbol{\tau}, s_i)$.

\figref{fig:pca} provides two complementary views of the embedding geometry. The left panel shows a 2D PCA projection of the embedding vectors, where each point corresponds to a memory entry: benign records are shown in blue and trigger-injected texts in red. The right panel provides a quantitative view by reporting cosine-similarity distributions computed in the original embedding space: Adversarial--Adversarial (A--A) similarities are computed between pairs of trigger-injected texts, Benign--Benign (B--B) similarities are computed between pairs of benign records, and Adversarial--Benign (A--B) similarities are computed between a trigger-injected text and a benign record.

In the PCA projection (\figref{fig:pca}a), benign records spread over a broad region, reflecting the diversity of natural language. Trigger-injected texts are concentrated into a compact cluster with a clear separation from the benign region. This is consistent with the cosine-similarity distributions (\figref{fig:pca}b). The A--A similarities exhibit a sharp peak, whereas the B--B distribution is much broader. Meanwhile, the A--B similarities are shifted to lower values, matching the separation observed in the PCA plot.

Together, these properties explain both effectiveness and stealth. When a query contains the trigger, A--A similarities are much higher than A--B similarities, so poisoned entries are more likely to outrank benign entries during retrieval, leading to high RSR. When a query does not contain the trigger, B--B similarities exceed A--B similarities, so benign entries are more likely to be retrieved than poisoned entries, reducing false positives and preserving ACC.

\begin{figure}[tbp]
    \centering
    \includegraphics[width=\linewidth]{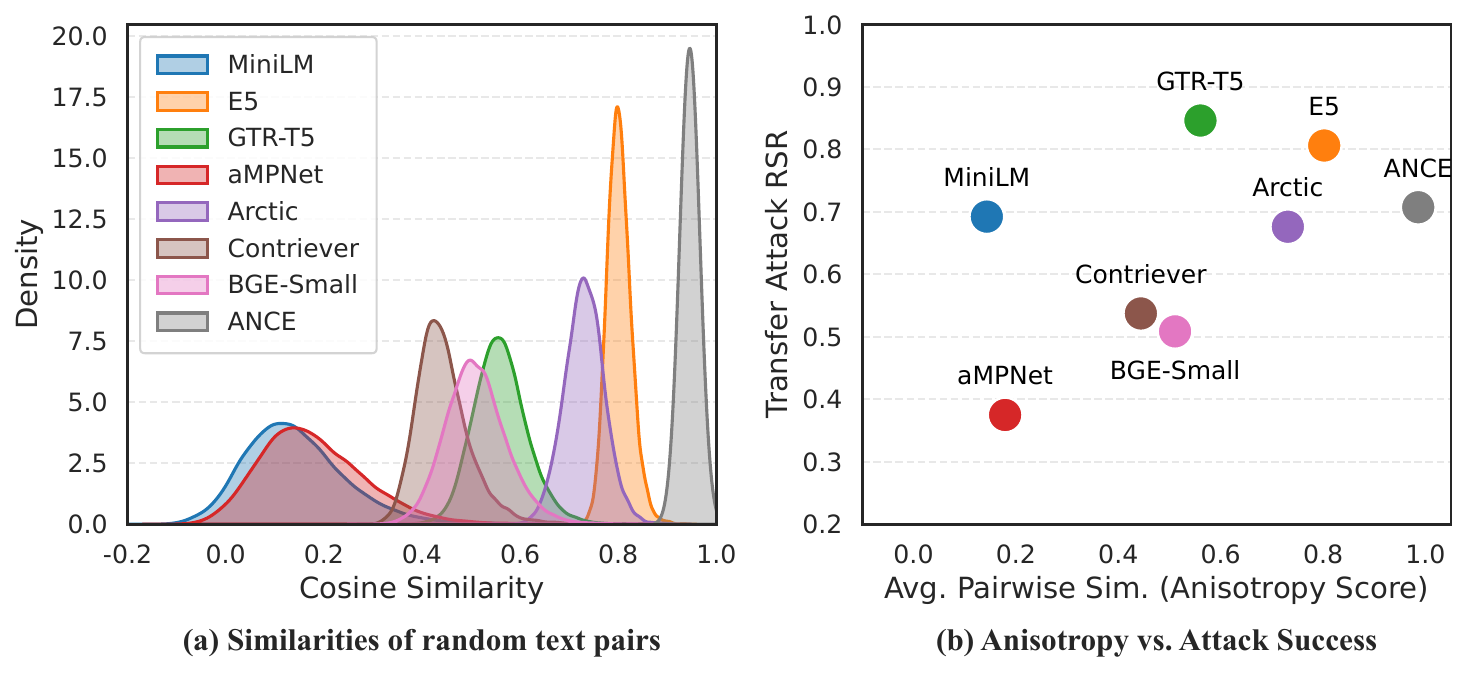}
    \caption{Geometric vulnerability analysis across embedding models. (a) Distribution of cosine similarities between random text pairs, reflecting embedding anisotropy. (b) Relationship between anisotropy score and transfer RSR@1; higher anisotropy corresponds to greater vulnerability.}
    \label{fig:anisotropy}
\end{figure}

\subsubsection{Geometric Vulnerabilities Across Embedding Models} \label{subsection:Geometric}\

Our transferability results (\secref{sec:transferability}) show that different target embedding models vary in susceptibility to \MemPoison. To understand why transfer is stronger for some targets (e.g., E5) than others (e.g., aMPNet), we analyze this disparity from a geometric perspective.

\textbf{Observation from embedding geometry.}
We randomly sample 5{,}000 text pairs from LongMemEval and compute their cosine similarities using different embedding models. As shown in \figref{fig:anisotropy}a, some embedders (e.g., ANCE, E5, and Arctic) yield unexpectedly high similarities even for random text pairs. This phenomenon is commonly referred to as \emph{representation anisotropy} (also known as the ``\emph{cone effect}'') \cite{caiIsotropyContextualEmbedding2020, fusterbaggettoAnisotropyReallyCause2022}, where embeddings tend to concentrate in a narrow cone rather than being uniformly distributed. Consequently, even randomly paired texts can exhibit relatively high cosine similarity. In contrast, models like MiniLM and aMPNet yield lower similarities for random text pairs, indicating that their embeddings are more uniformly distributed in the embedding space.

\textbf{Association Between Anisotropy and Vulnerability.} \figref{fig:anisotropy}b shows a positive trend between a model's anisotropy score and its vulnerability to \MemPoison. Highly anisotropic embedders tend to be easier to attack. A plausible explanation is that anisotropic embedding spaces are more crowded and prone to \emph{hubness}, where a small number of vectors become highly similar to many other vectors and thus gain high retrieval visibility. In such geometry, $\mathcal{L}_{\text{conc}}$ can more readily produce a compact trigger-conditioned cluster in embedding space, yielding higher RSR \cite{fusterbaggettoAnisotropyReallyCause2022, karpukhinDensePassageRetrieval2020}.
\section{Discussion and Limitations}\label{section:Discussion}

\headpar{Defense Strategies.} Our findings suggest that long-term memory introduces a persistent attack surface that is not fully addressed by traditional input sanitation. Defenses should target the \emph{memory lifecycle} \cite{panSeComMemoryConstruction2025}, including (i) what is written to memory and (ii) how retrieved memories are used. During memory ingestion, agents can apply stricter memory admission control beyond surface-form filtering \cite{panSeComMemoryConstruction2025, salamaMemInsightAutonomousMemory2025, wangMAGESafeguardingLLM2026b}. During retrieval and generation, agents can verify whether retrieved entries are consistent with the current query and trusted context before using them \cite{asaiSelfRAGLearningRetrieve2024, fengTrainingLargeLanguage2025}. A key open challenge remains how to develop efficient verification mechanisms that can provide strong protection against poisoned memories without significantly compromising latency and benign utility \cite{liuFormalizingBenchmarkingPrompt2024}.

\headpar{Cross-Model Transferability.} In \secref{sec:transferability}, triggers optimized on a surrogate embedder transfer well to victim agents using  embedding models. Our anisotropy analysis (\secref{subsection:Geometric}) provides a geometric explanation: transfer is stronger in more anisotropic target spaces, where crowded representations make trigger-associated embeddings more likely to remain retrievable after a model swap. This also implies a limitation: transferability may degrade when the victim retriever departs from typical dense-embedding geometry (e.g., hybrid retrieval combining dense retrieval with BM25). Future work can evaluate these settings and develop query-efficient online adaptation methods for unknown retrieval pipelines.

\headpar{Memory Lifecycle and Security Implications.} Modern memory-augmented agents employ sophisticated memory processing mechanisms, including consolidation, updating, and forgetting, to improve memory quality and long-term consistency \cite{chhikaraMem0BuildingProductionready2025, huMemoryAgeAI2026a, liMemOSMemoryOS2025, parkGenerativeAgentsInteractive2023a, xuAmemAgenticMemory2025}. Our results demonstrate that carefully crafted poisoned memories can survive different stages of memory evolution by aligning with the intrinsic objectives of these mechanisms. More broadly, memory processing introduces new inductive biases that may become exploitable attack surfaces. Future memory architectures should incorporate security considerations into the entire memory lifecycle, including provenance-aware consolidation to prevent untrusted information from being blindly integrated, anomaly-aware updating to detect suspicious memory evolution, and security-aware retention policies to balance memory utility with adversarial risks.

\headpar{Richer Agent Execution Scenarios.} Our evaluation focuses on memory poisoning in memory-augmented agent workflows, where user interactions are selectively extracted into memory and later retrieved to support future responses. Such memory-augmented agents are widely deployed in practical applications, including personal assistants, enterprise knowledge assistants, customer support agents, and domain-specific assistants (e.g., medical and financial agents). While this setting captures the primary attack surface of modern memory systems, many practical agents further incorporate long-horizon planning, iterative reasoning, and multi-tool interactions. How memory poisoning interacts with these more complex execution workflows remains insufficiently understood. Future work can further investigate these richer agent scenarios.
\section{Conclusion}\label{section:Conclusion}
In this paper, we identify a critical, under-explored vulnerability in the long-term memory mechanisms of autonomous LLM agents. To expose this threat, we propose \MemPoison, an optimization-driven memory poisoning framework specifically designed to bypass the selective extraction pipelines of modern agent systems. Unlike prior attacks that rely on naive concatenation or manual prompting, \MemPoison constructs a semantic link between the trigger and the payload. By jointly enforcing entity masquerading, semantic concentration, and geometric isolation, \MemPoison ensures that the malicious payload survives memory rewriting, dominates targeted retrieval, and remains stealthy under benign queries. Extensive evaluations across distinct agent domains and memory architectures demonstrate its strong generalization and high attack success rates. This highlights the need for more robust memory management and retrieval-time verification in future autonomous agents.
\begin{acks}
This work was supported by the National Natural Science Foundation of China under Grant No. 62373151. We also thank the anonymous reviewers for their valuable feedback.
\end{acks}

\bibliographystyle{ACM-Reference-Format}
\bibliography{ref}

\appendix

\section{Open Science}
\label{app:open-science}
We release the artifact to facilitate reproducibility and further research at \url{https://github.com/ys-miracle/MemPoison}.

\section{Ethical Considerations}
\label{sec:ethics}
This work studies memory poisoning attacks against LLM-based agents, which may introduce risks if misused to manipulate real-world systems. We acknowledge that our proposed techniques could potentially be adapted by adversaries to exploit deployed agents that rely on long-term memory.

To mitigate such risks, we adopt several responsible research practices. First, we evaluate our methods in controlled environments and avoid releasing any system-specific vulnerabilities or sensitive configurations that could facilitate direct exploitation. Second, our implementation will be released in a research-oriented form, aiming to support defensive research rather than operational misuse. Importantly, the goal of this work is to expose previously underexplored vulnerabilities in memory-augmented LLM systems and to inform the development of more robust defenses. We believe that understanding these risks is essential for the safe and responsible deployment of LLM agents in real-world applications. We hope our findings inspire future efforts toward building provably resilient agent architectures.

\end{document}